\documentclass[article,ij4uq]{ij4uq}              

\usepackage{subcaption}
\usepackage{url}
\usepackage{xcolor}
\usepackage{siunitx}
\usepackage{amssymb}
\usepackage{bbm}
\usepackage{graphicx}
\usepackage{amsthm}
\usepackage[]{algorithm2e}
\usepackage{mathtools}
\usepackage{makecell}
\usepackage{nicefrac}

\usepackage{glossaries}
\newacronym{fem}{FEM}{finite element method}
\newacronym{dof}{DoF}{degrees of freedom}
\newacronym{fe}{FE}{finite element}
\newacronym{nurbs}{NURBS}{non-uniform rational B-splines}
\newacronym{te}{TE}{transverse electric}
\newacronym{tm}{TM}{transverse magnetic}
\newacronym{pec}{PEC}{perfect electric conductor}
\newacronym{uq}{UQ}{uncertainty quantification}
\newacronym{pml}{PML}{perfectly matched layers}
\newacronym{mim}{MIM}{metal-insulator-metal}
\newacronym{rhs}{RHS}{right-hand side}
\newacronym{sqp}{SQP}{sequential quadratic programming}
\newacronym{gpc}{gPC}{generalized polynomial chaos}
\newacronym{hf}{HF}{high-frequency}
\newacronym{mc}{MC}{Monte Carlo}
\newacronym{pde}{PDE}{partial differential equation}
\newacronym{pdf}{PDF}{probability density function}
\newacronym[longplural={quantities of interest}]{qoi}{QoI}{quantity of interest}
\newacronym{rv}{RV}{random variable}
\makeglossaries

\DeclareMathOperator*{\argmax}{arg\,max}

\begin{document}

\title{Enhanced adaptive surrogate models with applications in uncertainty quantification for nanoplasmonics}
\titlehead{Enhanced adaptive surrogate models for uncertainty quantification}
\authorhead{N. Georg, D. Loukrezis, U. R\"omer \& S. Sch\"ops}
\corrauthor[1,2,3]{Niklas Georg}
\author[2,3]{Dimitrios Loukrezis}
\author[1]{Ulrich R\"omer}
\author[2,3]{Sebastian Sch\"ops}
\corremail{n.georg@tu-braunschweig.de}
\corraddress{Institut f{\"u}r Dynamik und Schwingungen, Technische Universit{\"a}t Braunschweig, Schleinitzstra{\ss}e 20, D-38106 Braunschweig, Germany}
\address[1]{Institut f{\"u}r Dynamik und Schwingungen, Technische Universit{\"a}t Braunschweig, Schleinitzstra{\ss}e 20, D-38106 Braunschweig, Germany}
\address[2]{Centre for Computational Engineering, Technische Universit{\"a}t Darmstadt, Dolivostra{\ss}e 15, D-64293 Darmstadt, Germany}
\address[3]{Institut f{\"u}r Teilchenbeschleunigung und Elektromagnetische Felder (TEMF), Technische Universit{\"a}t Darmstadt, Schlo{\ss}gartenstra{\ss}e 8, D-64289 Darmstadt, Germany}

\abstract{

We propose an efficient surrogate modeling technique for uncertainty quantification. The method is based on a well-known dimension-adaptive collocation scheme. We improve the scheme by enhancing sparse polynomial surrogates with conformal maps and adjoint error correction. The methodology is applied to Maxwell's source problem with random input data. This setting comprises many applications of current interest from computational nanoplasmonics, such as grating couplers or optical waveguides. Using a non-trivial benchmark model we show the benefits and drawbacks of using enhanced surrogate models through various numerical studies. 
The proposed strategy allows us to conduct a thorough uncertainty analysis, taking into account a moderately large number of random parameters. 

}


\maketitle

\section{Introduction}
\label{sec:introduction}

The numerical solution of \glspl{pde} with random input data has been receiving considerable attention in the last decades in the context of \gls{uq}. Numerical \gls{uq} methods are continuously improved to address large-scale problems with many input parameters, which still pose a computational challenge nowadays. The key property to reduce computational costs in high dimensions is a holomorphic dependency of the \gls{pde} solution on the input parameters. Such holomorphy results have been established for a variety of different problem classes \cite{chkifa2015breaking,scarabosio2016} and allow the use of spectral stochastic methods \cite{xiu2009, ghanem1991,  babuska2004, babuska2010, xiu2002, xiu2005} in combination with adaptive schemes, see, e.g., \cite{chkifa2014, gerstner2003, narayan2014, nobile2008a, ernst2018} for the case of stochastic collocation.

An alternative, sometimes complementary, approach for the numerical solution of parametric problems is model order reduction, see \cite{benner2015} and the references therein. Model order reduction based on moment-matching \cite{benner2015survey,bodendiek2014adaptive} can be used to derive a rational parametric approximation, which is appealing in the case of reduced parametric regularity. Rational Pad{\'e}-type approximations have recently been employed for a stochastic Helmholtz problem \cite{bonizzoni2018pade}. 
Moreover, in \cite{chantrasmi2009pade} a Pad{\'e}-Legendre method was introduced to cope with discontinuous response surfaces, where it was also noted that high-dimensional settings are still difficult to address. Another alternative are (multilevel) Monte Carlo methods, which can handle high-dimensional parameter spaces and more general settings with reduced smoothness. A recent analysis of such an approach for a Helmholtz transmission problem, with point evaluations as quantitites of interest, was presented in \cite{scarabosio2019multilevel}. The Helmholtz equation is recovered when two-dimensional versions of our model problem are considered. However, we quantify uncertainties in scattering parameters, which are more regular. The last class of methods which are mentioned here are perturbation methods \cite{silva2017, doelz2019}. A perturbation approach can lead to very efficient numerical methods, but is also not considered here, because the uncertainty in the input parameters of our models can be quite large.

The present study was motivated by the fact that the computational cost of constructing a sparse surrogate model  can still be quite high for various applications. In particular, even if sufficient smoothness is present to allow for sparse approximation, achieving a reasonable error level in practical applications may require collocation grids with many points. Hence, in this work, we improve a state of the art adaptive stochastic collocation method, based on dimension-adaptivity \cite{gerstner2003} and weighted Leja interpolation \cite{narayan2014}. To this end, we combine conformal maps and adjoint error estimation and correction. Conformal maps have been put forth in \cite{jantsch2018sparse,trefethen2013} for the acceleration of interpolation and quadrature methods, but have not received much attention in the \gls{uq} context so far. Adjoint error correction in turn was considered in \cite{jakeman2015,butler2013propagation} in the context of Clenshaw Curtis collocation and the stochastic Galerkin method. The combination of both methods in the context of uncertainty quantification has not been considered, to the best of our knowledge. The resulting collocation scheme is able to address a moderately high number of random model parameters. Moreover, weighted Leja nodes can handle almost arbitrary input probability distributions \cite{narayan2014, loukrezis2019assessing, loukrezis2019approximation, farcas2019multilevel, Bos2019, loukrezis2019adaptive} and are ideally suited for adaptivity \cite{chkifa2014}. In order to efficiently steer the adaptivity, we derive an adjoint representation of the stochastic error. Based on this error formula, the convergence order is enhanced through extrapolation. Finally, conformal maps offer the potential to further enhance the numerical accuracy by suitably transforming the required region of holomorphy.

We consider Maxwell's source problem as a model class which is relevant for a wide variety of applications. This model problem is particularly important in computational nanoplasmonics. Plasmonic structures offer great potential for subwavelength optics and optoelectronics \cite{genet2007} and have been intensively studied from both a fundamental and an application point of view in recent years. 
With the aforementioned \gls{uq} methods, studying stochastic parameter variations within the numerical simulation of plasmonic structures comes into reach. This is highly relevant, as relatively large variabilities can be observed, see, e.g. \cite{preiner2008}.
Although, not considered in the present work, the inverse \gls{uq} problem is also of high relevance, due to the intrinsic difficulty in measuring material dispersion properties. Instead, we focus on the propagation of uncertainties from the model inputs to the outputs. In particular the proposed framework allows to compute moments, probability distributions, failure probabilities, and global sensitivities for physical \glspl{qoi}. 
In similar physical settings, \gls{uq} studies have been conducted in the recent works  \cite{pitelet2019influence, schmitt2019optimization, loukrezis2019robust}, which, however, employ less advanced numerical methods. In \cite{weng2015} \gls{uq} for a silicon photonic device has been addressed, considering a low-dimensional correlated random input parameter vector. Additionally, in comparison to recent theoretical studies \cite{hiptmair2018, scarabosio2016}, we employ additional techniques for convergence acceleration and consider a more complex numerical example.

The rest of this paper is structured as follows. 
In Section~\ref{sec:uq} we describe the numerical method for enhanced surrogate modeling.
In Section~\ref{sec:maxwell} we introduce Maxwell's source problem, its finite element discretization, and its parametrization.
In Section~\ref{sec:aplication}, the developed method is used to conduct a \gls{uq} study for a non-trivial nanoplasmonics application, namely an optical grating coupler. In the last section we give some concluding remarks. \section{Enhanced Surrogate Modeling}
\label{sec:uq}
In this section, we consider the general parametric problem of finding
\begin{equation}
\label{eq:pdeoper}
\ensuremath{\mathbf{u}}(\ensuremath{\mathbf{y}}) \in V \text{ s.t.} \quad a_\ensuremath{\mathbf{y}}(\ensuremath{\mathbf{u}}(\ensuremath{\mathbf{y}}),\ensuremath{\mathbf{v}})  = l_\ensuremath{\mathbf{y}}(\ensuremath{\mathbf{v}}) \quad \forall \ensuremath{\mathbf{v}} \in V,
\end{equation}
where $V$ denotes a suitable Hilbert space and $\mathbf y\in \Xi \subset \mathbb R^N$ denotes the input parameter vector. Problem \eqref{eq:pdeoper} may represent the model of Section \ref{sec:parametrized_model}, or other parameterized differential equations with a continuous sesquilinear form  $a_\ensuremath{\mathbf{y}}(\cdot, \cdot)$ and a continuous (anti)linear form $l_\ensuremath{\mathbf{y}}(\cdot)$. Note that boldface letters are used to indicate matrices and vectors. Since the solutions governed by Maxwell's equations are typically vector-valued, this convention is also used for $\mathbf u$ and $\mathbf v$ in \eqref{eq:pdeoper}.
We assume the map $\ensuremath{\mathbf{u}}:\Xi \rightarrow V$ to be well-defined and smooth, which is often the case for parameterized differential equations, see e.g. \cite{babuska2007} for elliptic problems and \cite{chkifa2015breaking} for other types of \glspl{pde}.
We are interested in the model's response which may be the solution $\ensuremath{\mathbf{u}}(\ensuremath{\mathbf{y}})$ itself or a bounded linear functional $J_\ensuremath{\mathbf{y}}\left(\ensuremath{\mathbf{u}}\left(\mathbf{y}\right)\right)$, commonly referred to as the \gls{qoi}.
In this work, we focus on single-valued and complex \glspl{qoi}, i.e. $J_\ensuremath{\mathbf{y}}\left(\ensuremath{\mathbf{u}}\left(\mathbf{y}\right)\right) \in \mathbb{C}$.
For brevity of notation and owing to the well-posedness of the system, we shall replace $J_\ensuremath{\mathbf{y}}\left(\ensuremath{\mathbf{u}}\left(\mathbf{y}\right)\right)$ with $J\left(\mathbf{y}\right)$, where $J$ can be understood as an abstract representation of the map from the input parameters to the \gls{qoi}.

We now assume that the input parameters are given as independent \glspl{rv} $Y_n$, $n=1,2,\dots,N$.
We introduce the random vector $\mathbf{Y}=\left(Y_1, Y_2, \dots,Y_N\right)^\top$, defined on the probability space $\left(\Theta, \Sigma, P\right)$, where $\Theta$ denotes the sample space, $\Sigma$ the sigma-algebra of events and $P$ the probability measure, its image set $\Xi = \Xi_1 \times \Xi_2 \cdots \times \Xi_N \subset \mathbb{R}^N$ and its \gls{pdf} $\varrho\left(\mathbf{y}\right) = \prod_{n=1}^N \varrho_n\left(y_n\right)$, such that $\mathbf{Y}: \Theta \rightarrow \Xi$ and $\varrho:\Xi \rightarrow \mathbb{R}_+$. 
Then, the parameter vector represents a realization of the random vector, i.e. $\mathbf{y} = \mathbf{Y}\left(\theta\right) \in \Xi$, $\theta \in \Theta$. Assuming independence is necessary for the tensor-product constructions in the collocation method, however, dependence could also be taken into account through a suitable transformation, for instance, Rosenblatt or Nataf transformations \cite{jankoski2019stochastic,lebrun2009rosenblatt}. 
In view of this transformation, we assume in this section that the image set $\Xi$ is given as the hypercube $[-1,1]^N$, for simplicity.

Now, the \gls{qoi} is itself a \gls{rv} and we are interested in quantifying uncertainty, e.g. by computing its moments, \gls{pdf}, quantiles, etc.
In the case where the \gls{qoi} is smooth (ideally, analytic) with respect to the input \glspl{rv}, spectral \gls{uq} methods \cite{lemaitre2010, xiu2010} may be employed.
Then, $J$ is particularly well suited to be approximated by polynomials such that
\begin{equation}
\label{eq:mtermsapprox}
J\left(\mathbf{y}\right) \approx \widetilde{J}\left(\mathbf{y}\right) 
= \sum_{m=0}^{M} s_m \Psi_m\left(\mathbf{y}\right),
\end{equation}
where $\Psi_m:\Xi \rightarrow \mathbb R$ are multivariate polynomials and \mbox{$s_m \in \mathbb{C}$} the associated coefficients, and fast convergence can be expected.
Once an approximation in the form of \eqref{eq:mtermsapprox} is available, it can be used as an inexpensive substitute of the original computational model for sampling-based computations.
Alternatively, some statistical information regarding the \gls{qoi} can be derived directly from the coefficients.
In the context of the present work, we will use approximations in the form of \eqref{eq:mtermsapprox}
based on sparse grid interpolation \cite{babuska2010, barthelmann2000, bungartz2004, chkifa2014, klimke2005, narayan2014, nobile2008, schieche2012, xiu2005}, which can be combined with a conformal mapping.

\subsection{Univariate interpolation and conformal maps}
\label{subsec:univ_mapping}
We first discuss univariate interpolation in some detail, since it is also the key building block for the tensor product constructions used in the multivariate case. In particular, we consider the univariate function $f:[-1,1]\rightarrow \mathbb C,$
\begin{align}
&f(y)\coloneqq J(y,0,\ldots,0).
\end{align}

\begin{figure}
\centering
\includegraphics{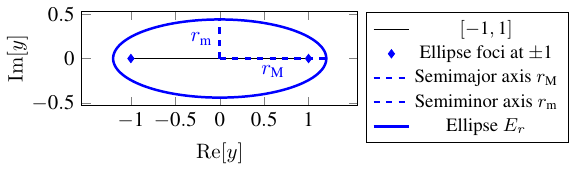}
\caption{Bernstein ellipse $E_r$ of size $r=r_{\text M}+r_\text{m}$.}
\label{fig:Bernstein}
\end{figure}

\begin{figure}
\begin{subfigure}[b]{0.47\textwidth}
\hspace{1em}\includegraphics{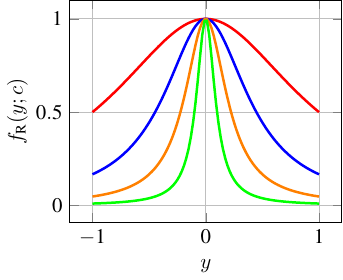}
\caption{}
\label{fig:runge}
\end{subfigure}\hfill
\begin{subfigure}[b]{0.47\textwidth}
\includegraphics{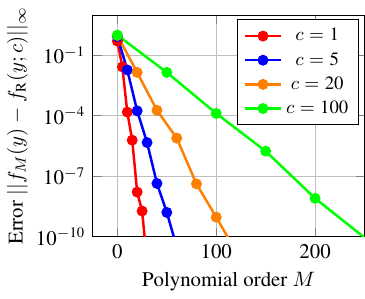}
\caption{}
\label{fig:conv_runge}
\end{subfigure}
\caption{a) Runge function $f_\text{R}(y;c)$ for different $c\in \mathbb R^+$ and $y\in[-1,1]$. b) Geometric convergence rates of Leja interpolants.}
\end{figure}

We assume that $f$ is analytic on $[-1,1]$ and can be analytically extended  onto $E_r$, where $E_r$ refers to an open Bernstein ellipse of size $r$, i.e. an ellipse in the complex plane with foci at $\pm 1$ and semi-minor and semi-major axis summing up to $r>1$, as illustrated in Fig.~\ref{fig:Bernstein}. 
Then, cf. \cite[Theorem 8.2]{trefethen2013}, the error of the univariate polynomial best approximation $f^*_{M}$ of degree $M$ can be estimated as \begin{equation}
\| f-f^*_{M}\|_\infty \le \frac {C_\text{B} r^{-M}}{r-1},
\label{eq:Trefethen1}
\end{equation}
where $\|\cdot\|_\infty$ denotes the supremum-norm on $[-1,1]$ and the constant $C_\text{B}>0$ depends on the uniform bound of the analytic continuation of $f$ in $E_r$.  
We consider a polynomial interpolant 
\begin{equation}
f_M(y) \coloneqq \sum_{i=0}^{M} f(y^{(i)}) l_i(y), 
\end{equation}
where $\{l_i\}_{i=0}^M$ and $\{y^{(i)}\}_{i=0}^M$ denote univariate Lagrange polynomials and a set of distinct nodes, respectively. 
There holds
\begin{align}
\|f-f_M\|_{\infty} &\le (1+\Delta_M) \|f-f_M^*\|_{\infty} \le (1+\Delta_M) \frac {C_\text{B} r^{-M}}{r-1}, \label{eq:poly_conv}
\end{align}
where 
\begin{equation}
\Delta_M\coloneqq \max_{y \in [-1,1]} \sum_{i=0}^{M} |l_i(y)| 
\end{equation}
denotes the Lebesgue constant. If $\Delta_M$ grows sub-exponentially, the polynomial interpolation converges uniformly (for analytic functions). 
However, the convergence rate depends on the regularity of the analytic continuation of $f$ in the complex plane. This is illustrated by considering the Runge function 
\begin{equation}
f_\text{R}(y;c) = \frac 1 {1+cy^2},\, \ c\in \mathbb R^+, y\in[-1,1],
\end{equation} 
which is shown in Fig.~\ref{fig:runge}, as a benchmark example.
This function is analytic on $[-1,1]$ but the analytic continuation has a complex conjugate pole pair at $y=\pm i\frac 1 {\sqrt{c}}$, limiting the size of the largest Bernstein ellipse where the function $f_\text{R}$ is analytic. Fig.~\ref{fig:conv_runge} demonstrates the effect on the convergence rate, where for increasing constants $c$, corresponding to a reduced size of the region of analyticity, a reduced convergence rate can be observed. The plot shows the convergence of the polynomial interpolant associated to unweighted Leja points in the empirical supremum-norm with a cross-validation sample of size $1000$. 

\begin{figure}
\centering
\includegraphics{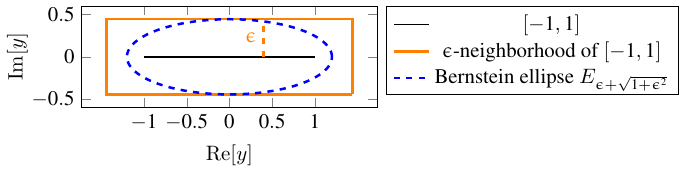}
\caption{$\epsilon$-neighborhood and largest interior Bernstein ellipse.}
\label{fig:neighborhood}
\end{figure}

\begin{figure}
\centering
\includegraphics{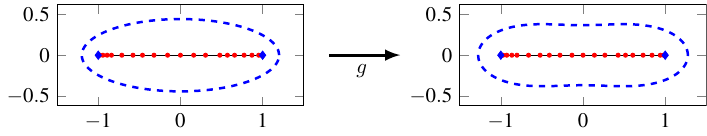}
\caption{Conformal map of a Bernstein ellipse.}
\label{fig:mapping}
\end{figure}

\begin{figure}
\centering
\includegraphics{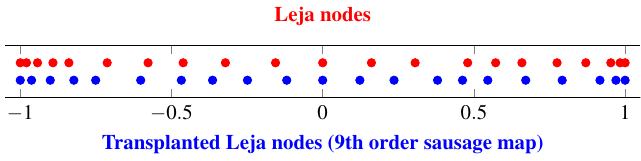}
\caption{Leja and transplanted Leja interpolation nodes.}
\label{fig:pointsdist}
\end{figure}

Hale and Trefethen \cite{hale2008} have raised and discussed the question \textit{whether polynomial methods are an optimal choice for functions analytic in an $\epsilon$-neighborhood}, in the context of numerical quadrature. Such a neighborhood is depicted in Fig.~\ref{fig:neighborhood} together with the largest Bernstein ellipse contained in its interior. They have pointed out that, in this case, superior methods to Gauss quadrature can be derived by conformally mapping the Bernstein ellipse $E_r$ to a \textit{straighter} region $\Omega_r=g(E_r)$, as illustrated in Fig.~\ref{fig:mapping}.
As will be discussed in the following, this approach is also beneficial for (polynomial) interpolation. 
In accordance with \cite{hale2008}, we focus in this work on conformal mappings $g:E_r\rightarrow \Omega_r$ which map the unit interval to itself, i.e. \begin{equation}g([-1,1])=[-1,1],\label{eq:map_cond1}\end{equation} and also fulfill \begin{equation}g(\pm 1) = \pm1\label{eq:map_cond2}.\end{equation} This ensures that the transplanted interpolation nodes \begin{equation}\{\hat y^{(i)}\}_{i=0}^M\coloneqq\{g\bigl(y^{(i)}\bigr)\bigr\}_{i=0}^M\end{equation} are still real numbers contained in the considered image set $\Xi_n$. There are various choices for $g$, see e.g. \cite{hale2009}, however, in this work we focus on the \textit{sausage mapping} proposed in \cite{hale2008}. It is defined by a $d$-th order Maclaurin expansion of the inverse sine function which is then normalized such that \eqref{eq:map_cond2} is fulfilled:
\begin{equation}\ensuremath{g_\mathrm{S}}(y;d)=\left(\sum_{i=0}^{\lfloor \nicefrac {(d-1)} 2 \rfloor}\frac {(2i)!}{4^i(2i+1)(i!)^2}\right)^{-1} \sum_{i=0}^{\lfloor \nicefrac {(d-1)} 2 \rfloor}\frac {(2i)!}{4^i(2i+1)(i!)^2}y^{2i+1}.\label{eq:sausage_map}
\end{equation}
An alternative mapping, due to Kosloff and Tal-Ezer \cite{kosloff1993}, is given by 
\begin{equation}
g_\text{KTE}(y;\alpha) = \frac{\arcsin (\alpha y)}{\arcsin{\alpha}},~\alpha\in(0,1). \label{eq:KTE}
\end{equation}
It can be observed that the transplanted nodes are more evenly distributed, see Fig.~\ref{fig:pointsdist}.

\begin{figure*}
\centering
\begin{subfigure}[b]{.48\textwidth}
\includegraphics{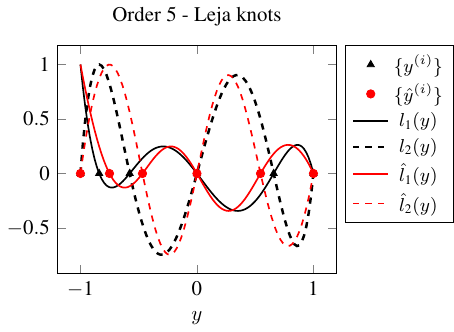}
\caption{Standard and mapped Lagrange polynomials.}
\label{fig:mappedpolys}
\end{subfigure}	\hfill
\begin{subfigure}[b]{.48\textwidth}
\includegraphics{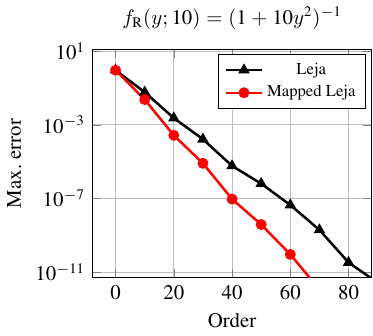}
\caption{Convergence of (mapped) interpolation of $f_\text{R}(y,10)$.}
\label{fig:Runge1DConv}
\end{subfigure}
\caption{Mapped Leja interpolation polynomials and numerical orders for increasing polynomial degrees.}
\end{figure*}

We interpolate the transplanted knots $\{\hat y^{(i)}\}$ using mapped Lagrange polynomials $\hat l_i = l_i \circ g^{-1}$, shown in Fig.~\ref{fig:mappedpolys}. Obviously, the mapped Lagrange polynomials also have the property 
\begin{equation}
\hat l_j(\hat y^{(i)}) = l_j \circ g^{-1}(\hat y^{(i)}) = l_j (y^{(i)}) =   \delta_{ij},
\end{equation}where $\delta_{ij}$ denotes the Kronecker delta. Thus, the mapped interpolant $\hat f_M$ is defined by
\begin{equation} \hat f_M(y)=\sum_{i=0}^M f\bigl(\hat y^{(i)}\bigr) \hat l_i(y).\end{equation}
To derive an error bound for the transplanted interpolation, we first introduce the function $h \coloneqq f\circ g$. We assume that $h$ can be continued analytically to $E_{\hat r}$, where it is uniformly bounded. Let $h_M$ be the $M$-th order polynomial interpolant of $h$ on the original nodes $\{y^{(i)}\}_{i=0}^M$. We observe that the mapped interpolant $\hat f_M$ is equivalent to $h_M \circ g^{-1}$ as
\begin{equation} 
\hat f_M= \sum_{i=0}^M f\left( g\left(y^{(i)} \right) \right) l_i \circ g^{-1}=h_M\circ g^{-1}.
\end{equation}
Due to \eqref{eq:map_cond1}, we obtain
\begin{align}||f-\hat f_M||_{\infty}  &= ||(f-\hat f_M) \circ g \circ g^{-1}||_{\infty}\\
&= ||(h-h_M) \circ g^{-1}||_{\infty} \\
&=||h-h_M||_{\infty} \\
&\le (1+\Delta_M)||h-h_M^*||_{\infty}\\
&\le (1+\Delta_M) \frac {\hat C_\text{B}\hat r^{-M}}{\hat r-1}. \label{eq:map_conv}
\end{align}
The convergence rate is improved if $\hat r>r$, which we confirm numerically in Fig.~\ref{fig:Runge1DConv}, where the $9$-th order sausage map $\ensuremath{g_\mathrm{S}}(y;9)$ is employed. Since $\ensuremath{g_\mathrm{S}}(y;9)^{-1}$ is not known analytically, we approximate the inverse mapping by a Chebyshev approximation of order 100 (up to machine precision). 

\begin{figure*}
\centering
\begin{subfigure}[b]{.46\textwidth}
\includegraphics{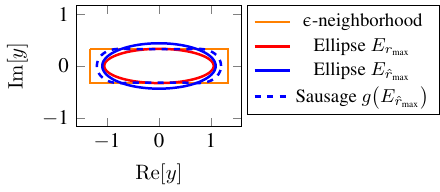}
\caption{Illustration of \textit{geometric gain estimation} for $\epsilon=0.3294$.}
\label{fig:sausage_in_box}
\end{subfigure}	\hfill
\begin{subfigure}[b]{.535\textwidth}
\hspace{1em}
\includegraphics{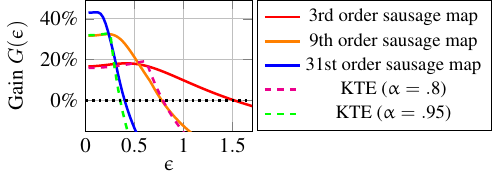}
\caption{Gain in convergence for different mappings.} 
\label{fig:neighborhood_gains}
\end{subfigure}
\caption{Convergence gain $G$ by employing mapped approximations.}
\end{figure*}

If the region of analyticity is known, one can estimate the gain of employing a conformal mapping a priori (based on the convergence estimates), as illustrated in Fig.~\ref{fig:sausage_in_box}. 
Let the size of the largest Bernstein ellipse in the region of analyticity be $r_\mathrm{max}$ and the size of the largest Bernstein ellipse, which is fully mapped into this region be $\hat{r}_\mathrm{max}$. 
The convergence of polynomial interpolation is then given by $\mathcal O\bigl((1+\Delta_M)\exp{(-\log(r_\mathrm{max})M)}\bigr)$ according to \eqref{eq:poly_conv} while the mapped interpolation converges as $\mathcal O\bigl((1+\Delta_M)\exp{(-\log(\hat{r}_\mathrm{max})M)}\bigr)$, see \eqref{eq:map_conv}. Assuming a sufficiently slowly growing Lebesgue constant $\Delta_M$, we consider the relative improvement in the asymptotic rate of geometric convergence, see \cite[Definition 6]{Boyd_2001aa}, given by 
\begin{equation}
G=\frac {\log \hat r_\mathrm{max}}{\log r_{\mathrm{max}}} - 1
\end{equation}
which can be attributed to the use of conformal maps.
We evaluate the gain $G$ for functions which are analytic in $\epsilon$-neighborhoods of $[-1,1]$, see Fig.~\ref{fig:neighborhood_gains}, by numerically computing $\hat r_\text{max}$ for different mappings. It should be noted that higher gains can be expected, if mappings would be employed, which are specifically tailored to the positions of the poles in the complex plane. However, usually the exact position of these poles is not known a priori and this approach is therefore not pursued any further. The interested reader is referred to \cite{hale2009}. In the remaining part of the paper, we work with the $9$-th order sausage mapping $\ensuremath{g_\mathrm{S}}(y;9)$ since a detailed comparison of different mappings is not in the scope of the present paper. The particular mapping is selected because it has already been established in \cite{hale2008, hale2009} and Fig.~\ref{fig:neighborhood_gains} confirms a significant gain in convergence for a substantial range of $\epsilon$-neighborhoods. Additionally, in contrast to the map \eqref{eq:KTE}, it does not introduce an artificial singularity.  

\subsection{Sparse grid interpolation}
\label{subsec:sparsegrid}
Approximations based on sparse grid interpolation are commonly referred to as sparse grid stochastic collocation methods \cite{babuska2010, nobile2008, xiu2005}. 
Those methods are based on combinations of univariate interpolation rules, defined by an interpolation level $\ell_n \in \mathbb{N}_0$, a monotonically increasing level-to-nodes function $m_n: \mathbb{N}_0 \rightarrow \mathbb{N}$, where $m_n\left(\ell_n\right) \eqqcolon m_{\ell_n}$ and $m_n\left(0\right) = 1$ and a grid of $m_{\ell_n}$ (mapped) interpolation nodes \begin{equation}Z_{\ell_n} = \Big\{\hat{y}_n^{(i_n)}\Big\}_{i_n=0}^{m_{\ell_n}-1}.\end{equation}
Introducing the multi-index $\boldsymbol{\ell} = \left(\ell_1,\ell_2,\dots,\ell_N\right) \in \mathbb{N}_0^N$, the tensor-product multivariate approximation is obtained as
\begin{equation}
J\left(\mathbf{y}\right) \approx \widetilde{J}\left(\mathbf{y}\right) = \sum_{\mathbf{i}:\hat{\mathbf{y}}^{(\mathbf{i})} \in Z_{\boldsymbol{\ell}}} J\left(\hat{\mathbf{y}}^{(\mathbf{i})}\right) \hat{L}_{\boldsymbol{\ell}, \mathbf{i}}\left(\mathbf{y}\right),
\end{equation}
where $\hat{\mathbf{y}}^{(\mathbf{i})} = \left(\hat{y}_1^{(i_1)}, \hat{y}_2^{(i_2)}, \dots, \hat{y}_N^{(i_N)}\right) \in Z_{\boldsymbol{\ell}}$ are multivariate interpolation nodes, uniquely identified by the multi-index $\mathbf{i} = \left(i_1, i_2, \dots, i_N\right) \in \mathbb{N}_0^N$ and $Z_{\boldsymbol{\ell}} = Z_{\ell_1} \times Z_{\ell_2} \times \cdots \times Z_{\ell_N}$ is the tensor grid of interpolation nodes. 

Moreover, $\hat{L}_{\boldsymbol{\ell}, \mathbf{i}}$ are mapped multivariate Lagrange polynomials, obtained by the composition $\hat{L}_{\boldsymbol{\ell}, \mathbf{i}} = L_{\boldsymbol{\ell}, \mathbf{i}} \circ \mathbf{g}^{-1}$ with
\begin{align}
\label{eq:lagrangePoly}
L_{\boldsymbol{\ell}, \mathbf{i}}\left(\mathbf{y}\right) = \prod_{n=1}^N l_{\ell_n, i_n}\left(y_n\right), 
\quad \text{where }
l_{\ell_n, i_n}(y_n) \coloneqq 
\begin{cases}
\prod_{k= 0, k \neq i_n}^{m_{\ell_n}-1} \frac{y_n-y_n^{(k)}}{y_n^{(i_n)} - y_n^{(k)}}, \quad &\ell_n \neq 0, \\
1, \quad &\ell_n=0.
\end{cases}
\end{align}
Obviously, for the trivial mapping $\mathbf g: \mathbf y\mapsto \mathbf y$, it holds $\hat{L}_{\boldsymbol{\ell}, \mathbf{i}}={L}_{\boldsymbol{\ell}, \mathbf{i}}$. 
More details on the multivariate coordinate-wise conformal mapping $\mathbf{g}$ will be given below. It should be noted that \eqref{eq:lagrangePoly} is used for the ease of exposition, in the actual implementation the barycentric representation should be used \cite{berrut2004}. 
Since $J\left(\mathbf{y}\right)$ has to be evaluated for each $\mathbf{y}^{(\mathbf{i})} \in Z_{\boldsymbol{\ell}}$, the complexity of the tensor-product approach is $\mathcal{O}\left(m_k^N\right)$, where \begin{equation}m_k := \max_{n} m_{\ell_n}.\end{equation}
This complexity can be mitigated to $\mathcal{O}\left(m_k \left(\log m_k\right)^{N-1}\right)$ by employing Smolyak sparse grids \cite{smolyak1963}, which typically result in an acceptable trade-off between approximation accuracy and complexity.
We introduce the  approximation level $k \in \mathbb{N}_0$ and define the multi-index set $\Lambda_{k}$, such that
\begin{equation}
\label{eq:smolyakset}
\Lambda_{k} \coloneqq \{\boldsymbol{\ell} \, : \, \left|\boldsymbol{\ell}\right|  = \ell_1 + \ell_2 + \dots + \ell_N \leq k\}.
\end{equation}
Then, the sparse grid of multivariate interpolation nodes $Z_{\Lambda_{k}}$ is constructed as
\begin{equation}
\label{eq:smolyakGrid}
Z_{\Lambda_{k}} = \bigcup_{k-N+1 \leq \left|\boldsymbol{\ell}\right|\leq k} Z_{\boldsymbol{\ell}},
\end{equation}
and the interpolation is given by 
\begin{equation}
\label{eq:smolyakApprox}
\mathcal{I}_{\Lambda_{k}}\left[J\right]\left(\mathbf{y}\right) = 
\sum_{\mathbf{i}: \hat{\mathbf{y}}^{(\mathbf{i})} \in Z_{\Lambda_{k}}} J\left(\hat{\mathbf{y}}^{(\mathbf{i})}\right) \hat{L}_{\boldsymbol{\ell},\mathbf{i}}\left(\mathbf{y}\right).
\end{equation}

\subsubsection{Mapped Leja nodes, hierarchical interpolation and adaptivity}
\label{subsubsec:adaptivity}
As shown in \cite{barthelmann2000}, Smolyak formulas are in general not interpolatory, unless based on nested sequences of univariate interpolation nodes, such that $Z_{\ell_n - 1} \subset Z_{\ell_n}$. Moreover, to ensure accuracy and fast convergence of the approximation, the interpolation nodes should be chosen in agreement with the \glspl{pdf} $\varrho_n\left(y_n\right)$.
We opt for weighted Leja interpolation nodes, as in \cite{narayan2014}. 
Omitting conformal mappings for the moment, we consider a univariate, continuous and positive weight function. 
Here, this weight function is given by a univariate \gls{pdf} $\varrho_n\left(y_n\right)$, $\varrho_n:\Xi_n \rightarrow \mathbb{R}_+$. 
A sequence of univariate Leja nodes $y_n^{(k)} \in \Xi_n$, $k=0,1,2,\dots$, can be constructed by solving the optimization problem
\begin{equation}
\label{eq:lejaopt}
y_n^{(K)} = \argmax_{y_n \in \Xi_n} \sqrt{\varrho_n\left(y_n\right)} \prod_{k=0}^{K-1}\left|y_n - y_n^{(k)}\right|,
\end{equation}
where the starting node $y_n^{(0)}$ is arbitrarily chosen.
For further details on the construction of weighted Leja nodes and an analysis of their properties, see \cite{narayan2014}.
We justify the choice of Leja nodes as follows.
First of all, Leja nodes satisfy the nestedness requirement by construction.
Secondly, they allow complete freedom in the choice of the level-to-nodes function $m_n\left(\ell_n\right)$.
Finally, they can be tailored to any given \gls{pdf}. 
In comparison, the commonly employed Clenshaw-Curtis nodes would restrict us to the rapidly growing level-to-nodes function $m_n\left(\ell_n\right) = 2^{\ell_n}+ 1$.
In the following, we employ the level-to-nodes function $m_n\left(\ell_n\right) = \ell_n + 1$, $\ell_n \in \mathbb{N}_0$, and denote with $y_n^{(\ell_n)}$ the single extra node corresponding to interpolation level $\ell_n$, i.e. $y_n^{(\ell_n)} = Z_{\ell_n} \setminus Z_{\ell_n -1}$.  
We also introduce for each parameter a conformal map $g_n$, as discussed in Section~\ref{subsec:univ_mapping}. 
Then, the mapped univariate Leja nodes $\hat y_n^{(k)}$ are obtained as $\hat y_n^{(k)} = g_n(y_n^{(k)})$. 
Of course, for the trivial map $g_n: y_n\mapsto y_n$ we recover the original Leja nodes $\hat y_n^{(k)} = y_n^{(k)}$. 
The multivariate mapping is then obtained as \begin{equation}\mathbf{g}(\mathbf{y}) = g_1(y_1) \cdots g_N(y_N).\end{equation} We note that the multivariate mapping $\mathbf g$ is conformal in each coordinate $y_n$.

In the multivariate case, nested grids of multivariate interpolation nodes can be constructed by enforcing the use of downward-closed (also, monotone or lower) multi-index sets \cite{chkifa2014, gerstner2003}.
Such sets are known to preserve the telescopic properties of the series in \eqref{eq:smolyakApprox} \cite{gerstner2003}.
Moreover, sequences of nested, downward-closed multi-index sets result in polynomial approximations of increasing accuracy \cite{chkifa2014}. 
Given a multi-index set $\Lambda$, let us first define its forward and backward neighbor multi-index sets, $\Lambda_+$ and $\Lambda_-$, respectively, such that
\begin{subequations}
\begin{align}
\Lambda_+ &\coloneqq \{\boldsymbol{\ell} + \mathbf{e}_n, \forall \boldsymbol{\ell} \in \Lambda, \forall n=1,\dots,N\}, \label{eq:forwardneighbors} \\
\Lambda_- &\coloneqq \{\boldsymbol{\ell} - \mathbf{e}_n, \forall \boldsymbol{\ell} \in \Lambda, \forall n=1,\dots,N \, : \, \ell_n>0\}, \label{eq:backwardneighbors}
\end{align}
\end{subequations}
where $\mathbf{e}_n$ is the $n$-th unit vector. 
Then, $\Lambda$ is said to be downward-closed if and only if
\begin{equation}
\label{eq:monotonicity}
\Lambda_- \subset \Lambda.
\end{equation}
Assuming now a multi-index $\boldsymbol{\ell} \notin \Lambda$ such that $\Lambda \cup \boldsymbol{\ell}$ is downward-closed, it holds that $Z_\Lambda \subset Z_{\Lambda \cup \boldsymbol{\ell}}$ and \begin{equation}\hat{\mathbf{y}}^{(\boldsymbol{\ell})} =  Z_{\Lambda \cup \boldsymbol{\ell}} \setminus Z_\Lambda,\end{equation}
where
\begin{equation}
Z_\Lambda = \bigcup_{\boldsymbol{\ell} \in \Lambda} Z_{\boldsymbol{\ell}}.
\end{equation}
Then, \eqref{eq:smolyakApprox} can be naturally transformed into the hierarchical interpolation
\begin{align}
\label{eq:hierarchinterpNd}
\mathcal{I}_{\Lambda \cup \boldsymbol{\ell}}\left[J\right]\left(\mathbf{y}\right)
= \mathcal{I}_{\Lambda}\left[J\right]\left(\mathbf{y}\right) + s_{\boldsymbol{\ell}} \, \hat{H}_{\boldsymbol{\ell}}\left({\mathbf{y}}\right),
\end{align}
where the coefficients $s_{\boldsymbol{\ell}} \in \mathbb{C}$, known as ``hierarchical surpluses'', are given by
\begin{equation}
\label{eq:HS}
s_{\boldsymbol{\ell}} = J\left(\hat{\mathbf{y}}^{(\boldsymbol{\ell})}\right) - \mathcal{I}_{\Lambda}\left[J\right]\left(\hat{\mathbf{y}}^{(\boldsymbol{\ell})}\right),
\end{equation}
and $\hat{H}_{\boldsymbol{\ell}}$ are multivariate mapped hierarchical polynomials, defined as
\begin{align}
\label{eq:hierarchPoly}
\hat{H}_{\boldsymbol{\ell}}\left(\mathbf{y}\right) = \prod_{n=1}^N \hat h_{\ell_n}\left(y_n\right), 
\quad\text{where }
\hat{h}_{\ell_n}\left(y_n\right) \coloneqq 
\begin{cases}
\prod_{k=0}^{\ell_n-1}\frac{g^{-1}_n(y_n) - y_n^{(k)}}{ y_n^{(\ell_n)} -  y_n^{(k)}}, \quad &\ell_n \neq 0, \\
1, \quad &\ell_n = 0.
\end{cases}
\end{align}
Again, by choosing $g_n$ as the identity map we recover standard hierarchical Lagrange polynomials. 

The use of (mapped) hierarchical polynomials has the advantage that the basis functions do not change as new nodes are added.
Moreover, the hierarchical surpluses $s_{\boldsymbol{\ell}}$ can be interpreted as error indicators, quantifying the contribution of the interpolation node $\hat{\mathbf{y}}^{(\boldsymbol{\ell})}$ to the already available approximation.
This interpretation motivates the adaptive construction of the sparse grid approximation based on a posteriori error estimates.
We consider a dimension-adaptive scheme, similar to the ones employed in \cite{chkifa2014, gerstner2003, klimke2005, narayan2014, schieche2012}, with minor modifications to address the case of complex \glspl{qoi}.
The scheme is presented in Algorithm \ref{algo:gensmolyak}.
A detailed description follows.

\begin{algorithm}[t]
\SetAlgoLined
\KwData{\gls{qoi} $J\left(\mathbf{y}\right)$, conformal map $\mathbf g$, multi-index \mbox{set $\Lambda$},  budget $B$}
\KwResult{sparse grid $Z_{\Lambda \cup \Lambda_+^{\mathrm{adm}}}$, \mbox{approximation $\mathcal{I}_{\Lambda \cup \Lambda_+^{\mathrm{adm}}}\left[J\right]$}}
\Repeat{simulation budget $B$ is reached}{
Compute the admissible set $\Lambda_+^{\mathrm{adm}}$, as in \eqref{eq:candidateset}. \\
Compute the hierarchical surpluses $s_{\boldsymbol{\ell}}$, $\forall \boldsymbol{\ell} \in \Lambda_+^{\mathrm{adm}}$, as in \eqref{eq:HS}. \\
Find the multi-index $\boldsymbol{\ell} \in \Lambda_+^{\mathrm{adm}}$ with the maximum error indicator $\left|s_{\boldsymbol{\ell}}\right|$. \\ 
Compute the approximation $\mathcal{I}_{\Lambda \cup \boldsymbol{\ell}}$, as in \eqref{eq:hierarchinterpNd}. \\
Set $\Lambda = \Lambda \cup \boldsymbol{\ell}$.}
\caption{Dimension-adaptive interpolation.} 
\label{algo:gensmolyak}
\end{algorithm}

Given a downward-closed multi-index set $\Lambda$, as well as the corresponding approximation $\mathcal{I}_\Lambda\left[J\right]$ and grid $Z_\Lambda$, we define the set of admissible neighbors $\Lambda_+^{\mathrm{adm}}$, such that
\begin{equation}
\label{eq:candidateset}
\Lambda_+^{\mathrm{adm}} \coloneqq \{\boldsymbol{\ell} \in \Lambda_+ \, : \, \boldsymbol{\ell} \notin \Lambda \:\:\: \text{and} \:\:\: \{\boldsymbol{\ell}\}_- \subset \Lambda \}.
\end{equation}
Expanding $\Lambda$ with admissible multi-indices $\boldsymbol{\ell} \in \Lambda_+^{\mathrm{adm}}$ guarantees that \eqref{eq:monotonicity} is satisfied, and we thus construct a sequence of nested downward-closed sets \cite{chkifa2014}.
In this work, the error indicator corresponding to each multi-index $\boldsymbol{\ell} \in \Lambda_+^{\mathrm{adm}}$ is chosen to be the modulus $\left| s_{\boldsymbol{\ell}} \right|$ of the corresponding complex hierarchical surplus, however, other choices are possible, e.g. $\max\left(|\mathrm{Re}\{s_{\boldsymbol{\ell}}\}|, |\mathrm{Im}\{s_{\boldsymbol{\ell}}\}|\right)$.
We update $\Lambda$ with the multi-index $\boldsymbol{\ell} \in \Lambda_+^{\mathrm{adm}}$ corresponding to the maximum error indicator $\left| s_{\boldsymbol{\ell}} \right|$.
The grid of interpolation nodes $Z_{\Lambda}$ and the approximation $\mathcal{I}_{\Lambda}$ are updated accordingly.
This procedure is continued iteratively, until a budget of model evaluations $B$ is reached. This criterion can be formulated as
\begin{equation}
\label{eq:exitconditions}
\#  Z_{\Lambda \cup \Lambda_+^{\mathrm{adm}}}  \geq B, 
\end{equation}
where $\#$ denotes the cardinality of a set.
If an approximation is not readily available, the algorithm is initiated with \mbox{$\Lambda = \{ \left(0,0,\dots,0\right)\}$}.
After the termination of the algorithm, the approximation is constructed using the set $\Lambda \cup \Lambda_+^{\mathrm{adm}}$.

\subsection{Adjoint error estimation and adaptivity} 
\label{sec:adjointAdapt}
We aim to improve Algorithm \ref{algo:gensmolyak} by using an adjoint error indicator to steer adaptivity.
Adjoint error estimation is well established in the context of the \gls{fem}, see \cite{becker2001optimal} and the references therein. 
It has been considered in a stochastic/parametric context \cite{butler2012posteriori,butler2013propagation, Roemer2015},
as well as for Clenshaw-Curtis adaptivity \cite{jakeman2015,schieche2012}. 
Due to the exponential growth of Clenshaw-Curtis nodes, adjoint error estimation can result in a significant reduction of computational cost. 
In this work, we demonstrate that adjoint techniques can be beneficial for Leja adaptivity, too.

In this section we rely on the fact that $J(\ensuremath{\mathbf{y}}) = J_{\ensuremath{\mathbf{y}}}(\ensuremath{\mathbf{u}}(\ensuremath{\mathbf{y}}))$, $J_{\ensuremath{\mathbf{y}}} : V \rightarrow \mathbb{C}$, is a linear functional with respect to $\ensuremath{\mathbf{u}}(\ensuremath{\mathbf{y}})$. 
However, generalizations to non-linear functionals are also possible, as in \cite[Chapter 3.2]{teckentrup2013}.
We rewrite the primal problem \eqref{eq:pdeoper} as an operator equation: $\forall \ensuremath{\mathbf{y}} \in \Xi$, find $\ensuremath{\mathbf{u}}(\ensuremath{\mathbf{y}}) \in V$, such that
\begin{align}
\langle L_{\ensuremath{\mathbf{y}}} \ensuremath{\mathbf{u}}(\ensuremath{\mathbf{y}}), \ensuremath{\mathbf{v}} \rangle = a_{\ensuremath{\mathbf{y}}}(\ensuremath{\mathbf{u}}(\ensuremath{\mathbf{y}}),\ensuremath{\mathbf{v}})  = l_\ensuremath{\mathbf{y}}(\ensuremath{\mathbf{v}})~~\forall \ensuremath{\mathbf{v}}\in V, \label{eq:primal_problem}
\end{align}
where $L_{\ensuremath{\mathbf{y}}}: V\rightarrow V^*$ denotes the primal operator and $V^*$ the dual space to $V$. The dual problem is given as: for all $\ensuremath{\mathbf{y}} \in \Xi$, find $\ensuremath{\mathbf{z}}(\ensuremath{\mathbf{y}}) \in V$, such that
\begin{align}
\langle\ensuremath{\mathbf{w}}, L_{\ensuremath{\mathbf{y}}}^* \ensuremath{\mathbf{z}}(\ensuremath{\mathbf{y}})\rangle = a_{\ensuremath{\mathbf{y}}}(\ensuremath{\mathbf{w}}, \ensuremath{\mathbf{z}}(\ensuremath{\mathbf{y}}))  = J_\ensuremath{\mathbf{y}}(\ensuremath{\mathbf{w}})~~\forall \ensuremath{\mathbf{w}}\in V,\label{eq:dual_problem}
\end{align}
where $L_{\ensuremath{\mathbf{y}}}^*: V\rightarrow V^*$ denotes the adjoint operator defined by 
\begin{align}
\langle L_{\ensuremath{\mathbf{y}}} \ensuremath{\mathbf{u}}, \ensuremath{\mathbf{v}} \rangle = \langle\ensuremath{\mathbf{u}}, L_{\ensuremath{\mathbf{y}}}^* \ensuremath{\mathbf{v}}\rangle ~~ \forall \ensuremath{\mathbf{u}}, \ensuremath{\mathbf{v}} \in V, \ \forall \ensuremath{\mathbf{y}} \in \Xi. 
\end{align}
The so-called primal-dual equivalence 
{
\begin{align}
J_\ensuremath{\mathbf{y}}(\ensuremath{\mathbf{u}}(\ensuremath{\mathbf{y}})) = \langle\ensuremath{\mathbf{u}}(\ensuremath{\mathbf{y}}), L_{\ensuremath{\mathbf{y}}}^*\ensuremath{\mathbf{z}}(\ensuremath{\mathbf{y}})\rangle = \langle L_{\ensuremath{\mathbf{y}}}\ensuremath{\mathbf{u}}(\ensuremath{\mathbf{y}}), \ensuremath{\mathbf{z}}(\ensuremath{\mathbf{y}}) \rangle = l_\ensuremath{\mathbf{y}}(\ensuremath{\mathbf{z}}(\ensuremath{\mathbf{y}}))
\end{align}}follows directly from these definitions. 
Given (mapped) polynomial approximations $\tilde{\ensuremath{\mathbf{u}}}, \tilde{\ensuremath{\mathbf{z}}}$ of the mappings $\ensuremath{\mathbf{u}}, \ensuremath{\mathbf{z}}: \Xi \rightarrow V$, we are interested in the error 
\begin{align}
\eta(\ensuremath{\mathbf{y}}) = J_\ensuremath{\mathbf{y}}\bigl(\ensuremath{\mathbf{u}}(\ensuremath{\mathbf{y}})-\tilde{\ensuremath{\mathbf{u}}}(\ensuremath{\mathbf{y}})\bigr) = a_\ensuremath{\mathbf{y}}\bigl(\ensuremath{\mathbf{u}}(\ensuremath{\mathbf{y}})-\tilde{\ensuremath{\mathbf{u}}}(\ensuremath{\mathbf{y}}), \ensuremath{\mathbf{z}}(\ensuremath{\mathbf{y}})\bigr) = l_\ensuremath{\mathbf{y}}\bigl(\ensuremath{\mathbf{z}}(\ensuremath{\mathbf{y}})\bigr) - a_\ensuremath{\mathbf{y}}\bigl(\tilde{\ensuremath{\mathbf{u}}}(\ensuremath{\mathbf{y}}), \ensuremath{\mathbf{z}}(\ensuremath{\mathbf{y}})\bigr).
\label{eq:adjoint_error}
\end{align}
Even if $\tilde{\ensuremath{\mathbf{u}}},\tilde{\ensuremath{\mathbf{z}}}$ are replaced by their finite element counterparts, the error according to \eqref{eq:adjoint_error} is not readily computable, as it would require the computation of the adjoint $\ensuremath{\mathbf{z}}$ for all $\ensuremath{\mathbf{y}} \in \Xi$. Following \cite{butler2012posteriori,butler2013propagation}, we propose to use the error indicator
\begin{align}
\tilde \eta(\ensuremath{\mathbf{y}}) = a_\ensuremath{\mathbf{y}}\bigl(\ensuremath{\mathbf{u}}(\ensuremath{\mathbf{y}})-\tilde{\ensuremath{\mathbf{u}}}(\ensuremath{\mathbf{y}}), \tilde{\ensuremath{\mathbf{z}}}(\ensuremath{\mathbf{y}})\bigr)= l_\ensuremath{\mathbf{y}}\bigl(\tilde{\ensuremath{\mathbf{z}}}(\ensuremath{\mathbf{y}})\bigr) - a_\ensuremath{\mathbf{y}}\bigl(\tilde{\ensuremath{\mathbf{u}}}(\ensuremath{\mathbf{y}}), \tilde{\ensuremath{\mathbf{z}}}(\ensuremath{\mathbf{y}})\bigr). \label{eq:errorEstContinuous}
\end{align}
By exploiting the continuity of the sesquilinearform $a_\ensuremath{\mathbf{y}}(\cdot, \cdot)$, it can be shown that the error indicator \eqref{eq:errorEstContinuous} converges faster than the mapped polynomial approximations  $\tilde{\ensuremath{\mathbf{u}}}, \tilde{\ensuremath{\mathbf{z}}}$
{
\begin{align}
\vert \eta(\ensuremath{\mathbf{y}})-\tilde \eta(\ensuremath{\mathbf{y}})\vert 
= \vert  a_\ensuremath{\mathbf{y}}\bigl(\ensuremath{\mathbf{u}}(\ensuremath{\mathbf{y}})-\tilde{\ensuremath{\mathbf{u}}}(\ensuremath{\mathbf{y}}), \ensuremath{\mathbf{z}}(\ensuremath{\mathbf{y}})-\tilde{\ensuremath{\mathbf{z}}}(\ensuremath{\mathbf{y}})\bigr)\vert\le C \| \ensuremath{\mathbf{u}}(\ensuremath{\mathbf{y}})-\tilde{\ensuremath{\mathbf{u}}}(\ensuremath{\mathbf{y}})\|_V  \| \ensuremath{\mathbf{z}}(\ensuremath{\mathbf{y}})-\tilde{\ensuremath{\mathbf{z}}}(\ensuremath{\mathbf{y}})\|_V. \label{eq:adjoint_conv}
\end{align}}In particular, considering for the moment the univariate case $N=1$, for simplicity, and assuming that $\mathbf{u},\mathbf{z}$ can be extended analytically onto open Bernstein ellipses $E_{\hat r_u},E_{\hat r_z}$, respectively and that there exists uniform bounds on their extensions. Then, we obtain 
\begin{align}
\| \eta - \tilde \eta \|_{\infty} \leq C_1 (1 +\Delta_M)^2 \frac{C_2 (\hat r_u \hat r_z)^{-M}}{(1-\hat r_u)(1-\hat r_z)},
\end{align} for $M$-point approximations of both $\ensuremath{\mathbf{u}}$ and $\ensuremath{\mathbf{z}}$. Hence, for \mbox{$\hat r_u = \hat r_z$}, $\tilde \eta$ exhibits twice the rate of geometric convergence.

We proceed by discussing the necessary adaptations to Algorithm \ref{algo:gensmolyak}, in order to incorporate the adjoint error indicator \eqref{eq:errorEstContinuous}.
Additionally to the (mapped) polynomial approximation \eqref{eq:hierarchinterpNd}  of the single-valued and complex \gls{qoi}, one needs to create (mapped) polynomial approximations $\tilde{\ensuremath{\mathbf{u}}}(\ensuremath{\mathbf{y}}), \tilde{\ensuremath{\mathbf{z}}}(\ensuremath{\mathbf{y}})$ of the primal and dual solution. 
The approximations are constructed with the same multi-index set $\Lambda$ as for the \gls{qoi}, using the same mapped polynomials $\hat{H}_{\boldsymbol{\ell}}(\ensuremath{\mathbf{y}})$.

Following \cite{jakeman2015}, we carry out the algorithmic modifications in the dimension-adaptive scheme. 
While Algorithm \ref{algo:gensmolyak} uses the error indicators $\left|s_{\boldsymbol{\ell}}\right|$, $\forall \boldsymbol{\ell} \in \Lambda_+^{\mathrm{adm}}$, by solving the respective linear problem, we suggest the use of the adjoint-based error indicators
$\left| \tilde{s}_{\boldsymbol{\ell}} \right|$, where $\tilde s_{\boldsymbol{\ell}} = \tilde{\eta}\bigl(\hat{\mathbf{y}}^{(\boldsymbol{\ell})}\bigr)$.
As before, we choose the multi-index with the maximum error indicator, solve the corresponding linear system and update the approximations of the primal and the dual solution, as well as of the \gls{qoi}. 
This scheme is summarized in Algorithm \ref{algo:gensmolyak2}. After the termination of the algorithm, the approximation can be constructed with the set $\Lambda \cup \Lambda_+^{\mathrm{adm}}$, such that the already computed adjoint-based error indicators are used as the hierarchical surpluses corresponding to the admissible neighbors, i.e.  $s_{\boldsymbol{\ell}}=\tilde{s}_{\boldsymbol{\ell}}, ~\forall \boldsymbol{\ell}\in \Lambda_+^{\mathrm{adm}}$. The error indicator \eqref{eq:errorEstContinuous} can be further employed in order to improve the (mapped) polynomial surrogate model of the \gls{qoi}. In particular, one can replace the single-valued \gls{qoi} $J(\ensuremath{\mathbf{y}})$ by 
\begin{align}\tilde J(\ensuremath{\mathbf{y}}) = \mathcal{I}_{\Lambda}[J](\ensuremath{\mathbf{y}}) +\tilde \eta(\ensuremath{\mathbf{y}}), \label{eq:approx_corrected}\end{align}
such that the computed mapped polynomial approximation is corrected by the adjoint-error indicator, before continuing with further approximation refinements using Algorithm \ref{algo:gensmolyak}. We emphasize that no additional linear equation system has to be solved in order to evaluate \eqref{eq:approx_corrected}.

\begin{algorithm}[t]
\SetAlgoLined
\KwData{$\mathbf g$, $\Lambda$, $B$ and $a_\mathbf{\mathbf y}$, $l_{\mathbf y}$, $J_{\mathbf y}$ as defined in \eqref{eq:primal_problem}-\eqref{eq:dual_problem}}
\KwResult{sparse grid $Z_{\Lambda \cup \Lambda_+^{\mathrm{adm}}}$, \mbox{approximation $\mathcal{I}_{\Lambda \cup \Lambda_+^{\mathrm{adm}}}\left[J\right]$}}
\Repeat{stopping criterion fulfilled}{
Compute the admissible set $\Lambda_+^{\mathrm{adm}}$, as in \eqref{eq:candidateset}. \\
Compute the error indicators $\left| \tilde{s}_{\boldsymbol{\ell}} \right|$, where $\tilde s_{\boldsymbol{\ell}} = \tilde{\eta}\left(\mathbf{y}^{(\boldsymbol{\ell})}\right)$, $\forall \boldsymbol{\ell} \in \Lambda_+^{\mathrm{adm}}$.\\  
Find the multi-index $\boldsymbol{\ell} \in \Lambda_+^{\mathrm{adm}}$ with the maximum error indicator. \\ 
Compute the hierarchical surpluses $s_{\boldsymbol{\ell}}$, $\ensuremath{\mathbf{u}}_{\boldsymbol{\ell}}$, $\ensuremath{\mathbf{z}}_{\boldsymbol{\ell}}$ as in \eqref{eq:HS}, by solving the linear problems for primal and dual solution.\\
Compute the approximation $\mathcal{I}_{\Lambda \cup \boldsymbol{\ell}}$, as in \eqref{eq:hierarchinterpNd}, and the corresponding approximations of primal and dual solution. \\
Set $\Lambda = \Lambda \cup \boldsymbol{\ell}$.}
\caption{Adjoint error-based, dimension-adaptive interpolation.} 
\label{algo:gensmolyak2}
\end{algorithm}
\section{Maxwell's Source Problem}
\label{sec:maxwell}

In the following, we introduce the model problem, i.e., Maxwell's source problem with periodic boundary conditions. Such a model can be used for instance to describe the coupling into a plasmonic grating coupler, which will be considered in Section~\ref{sec:aplication}.
We also introduce the \gls{fe} approximation and a parametric version of the model.

\subsection{Deterministic Problem}
We start with the time-harmonic Maxwell's equations,
\begin{subequations}
\begin{align}
\ensuremath{\nabla \times \ensuremath{\mathbf{E}}}  &=-j\omega \mu \ensuremath{\mathbf{H}} && \text{in }\ensuremath{D}, \\
\ensuremath{\nabla \times \ensuremath{\mathbf{H}}}  &= \ensuremath{\mathbf{J}}_\textrm{s}  + j\omega \varepsilon \ensuremath{\mathbf{E}}  && \text{in }\ensuremath{D},\\
\ensuremath{\nabla \cdot \left(\varepsilon \ensuremath{\mathbf{E}}\right)}   &=\rho  && \text{in }\ensuremath{D},\\
\ensuremath{\nabla \cdot \left(\mu \ensuremath{\mathbf{H}}\right)}   &=0  && \text{in }\ensuremath{D},
\end{align}
\end{subequations}
where $\ensuremath{\mathbf{E}}$ denotes the electric field phasor, $\ensuremath{\mathbf{H}}$ the magnetic field phasor, $\ensuremath{\mathbf{J}}_{\textrm{s}}$ the source current phasor, $\rho$ the charge density phasor, $\omega$ the angular frequency, $\varepsilon$ the dispersive complex-valued permittivity, $\mu$ the permeability and $\ensuremath{D}$ the computational domain to be specified. The permeability $\mu=\mu_0\mu_r$, where $\mu_{\mathrm{r}}$ and $\mu_0$ represent the relative and vacuum permeability, respectively, is assumed to be nondispersive. 
In absence of charges and source currents, i.e. $\rho=0$ and $\ensuremath{\mathbf{J}}_\textrm{s}=0$, the so-called $\text{curl}$-$\text{curl}$ equation reads
\begin{equation}
\nabla \times \left(\mu_{\mathrm{r}}^{-1} \nabla \times \ensuremath{\mathbf{E}}\right)- \omega^2 \varepsilon\mu_0 \ensuremath{\mathbf{E}} = 0 \qquad \text{in }\ensuremath{D}, \label{eq:curlcurl}
\end{equation}
to be endowed with appropriate boundary conditions. 

\begin{figure}
\centering	 
\includegraphics{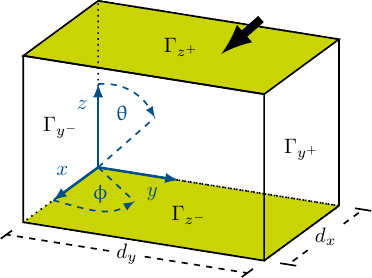}
\caption{Sketch of a unit cell representing the computational domain $\ensuremath{D}$. The black arrow indicates the incident wavevector $\ensuremath{\ensuremath{\mathbf{k}}^\mathrm{inc}}$.}
\label{fig:UnitCell}
\end{figure}

Given an infinitely periodic structure and a periodic excitation, the computational domain $\ensuremath{D}$ can be confined to a single unit cell of the periodic structure, based on Floquet's Theorem \cite[Chapter 13]{jin2015}. 
The unit cell is illustrated in Fig.~\ref{fig:UnitCell}.
Without loss of generality we assume periodicity in the $x$ and $y$ directions, whereas $\ensuremath{{\Gamma_{z^+}}}$ and $\ensuremath{{\Gamma_{z^-}}}$ denote the boundaries in the non-periodic direction.
At $\ensuremath{{\Gamma_{z^+}}}$ the structure is excited by an incident plane wave 
\begin{align}
\ensuremath{\ensuremath{\mathbf{E}}^\mathrm{inc}} = \ensuremath{\mathbf{E}}_0 e^{-j\ensuremath{\ensuremath{\mathbf{k}}^\mathrm{inc}} \cdot \ensuremath{\mathbf{r}}}, && \ensuremath{\ensuremath{\mathbf{k}}^\mathrm{inc}} = \begin{bmatrix} \ensuremath{k^{\mathrm{inc}}_x}\\\ensuremath{k^{\mathrm{inc}}_y}\\\ensuremath{k^{\mathrm{inc}}_z}\end{bmatrix} = -k_0
\begin{bmatrix} \sin \ensuremath{\theta^\mathrm{inc}} \cos \ensuremath{\phi^\mathrm{inc}}\\\ \sin \ensuremath{\theta^\mathrm{inc}} \sin \ensuremath{\phi^\mathrm{inc}}\\\ \cos\ensuremath{\theta^\mathrm{inc}}\end{bmatrix},
\end{align}
where $\ensuremath{\theta^\mathrm{inc}},~\ensuremath{\phi^\mathrm{inc}}$ are the angles of incidence and $k_0=\omega \sqrt{\ensuremath{\mu_0}\ensuremath{\varepsilon_0}}$ the wavenumber in vacuum. It is worth noting that, due to the oblique angles, the periodicity of the excitation differs from the geometrical periodicity of the structure. According to Floquet's theorem, we need to enforce periodic phase-shift boundary conditions $\text{on}~{\ensuremath{{\Gamma_{x^+}}}\cup \ensuremath{{\Gamma_{x^-}}}}$ and $\text{on}~{\ensuremath{{\Gamma_{y^+}}}\cup \ensuremath{{\Gamma_{y^-}}}}$, i.e.
\begin{subequations}
\begin{align}
\ensuremath{\mathbf{E}}\vert_{\ensuremath{{\Gamma_{x^+}}}} &= \ensuremath{\mathbf{E}}\vert_{\ensuremath{{\Gamma_{x^-}}}} e^{j\psi_x}, ~~~ \psi_x = -k^{\mathrm{inc}}_x d_x  \label{eq:phaseshiftBCx}\\
\ensuremath{\mathbf{E}}\vert_{\ensuremath{{\Gamma_{y^+}}}} &= \ensuremath{\mathbf{E}}\vert_{\ensuremath{{\Gamma_{y^-}}}} e^{j\psi_y}, ~~~ \psi_y = -k^{\mathrm{inc}}_y  d_y \label{eq:phaseshiftBCy}
\end{align}
\end{subequations}
where the phase-shifts $\psi_x,~\psi_y$ depend only on the wavevector $\ensuremath{\ensuremath{\mathbf{k}}^\mathrm{inc}}$ of the incident wave at $\ensuremath{{\Gamma_{z^+}}}$ and on the dimensions $\ensuremath{{d_x}},~\ensuremath{d_y}$ of the unit cell. 

To truncate the structure in the non-periodic direction at $\ensuremath{{\Gamma_{z^+}}}$, we employ a Floquet absorbing boundary condition \cite{jin2015} as derived in \ref{sec:AppendixA}. At $\ensuremath{{\Gamma_{z^-}}}$, a \gls{pec} boundary condition is applied to truncate the structure, 
however, different boundary conditions are also possible, e.g. again a Floquet absorbing boundary condition or \gls{pml} \cite{jin2015}. 
In summary, we are concerned with the boundary value problem
\begin{subequations}
\begin{align}
\nabla \times \left(\mu_{\mathrm{r}}^{-1} \nabla \times \ensuremath{\mathbf{E}}\right)- \omega^2 \varepsilon\mu_0 \ensuremath{\mathbf{E}} &= 0 && \text{in }\ensuremath{D}, \label{eq:MaxwellBVP_vol}\\
\ensuremath{\mathbf{E}}\vert_{\ensuremath{{\Gamma_{x^+}}}}e^{-j\psi_x} &= \ensuremath{\mathbf{E}}\vert_{\ensuremath{{\Gamma_{x^-}}}}  &&\text{on}~{\ensuremath{{\Gamma_{x^+}}}\cup \ensuremath{{\Gamma_{x^-}}}},\label{eq:MaxwellBVP_P1} \\
\ensuremath{\mathbf{E}}\vert_{\ensuremath{{\Gamma_{y^+}}}}e^{-j\psi_y} &= \ensuremath{\mathbf{E}}\vert_{\ensuremath{{\Gamma_{y^-}}}}  &&\text{on}~{\ensuremath{{\Gamma_{y^+}}}\cup \ensuremath{{\Gamma_{y^-}}}},\label{eq:MaxwellBVP_P2}\\
\ensuremath{\mathbf{n}} \times \ensuremath{\mathbf{E}} &= 0 && \text{on }\ensuremath{{\Gamma_{z^-}}}, \label{eq:MaxwellBVP_pec}\\
\ensuremath{\mathbf{n}} \times  \ensuremath{\mathbf{H}} +  \boldsymbol{\mathcal G}(\ensuremath{\mathbf{E}}) &= \boldsymbol{\mathcal F}^\text{inc} && \text{on } \ensuremath{{\Gamma_{z^+}}},\label{eq:MaxwellBVP_Floquet}
\end{align}\label{eq:MaxwellBVP}\end{subequations}
where $\boldsymbol{\mathcal G}(\ensuremath{\mathbf{E}})$ and $\boldsymbol{\mathcal F}^\text{inc}$ are derived and defined in \ref{sec:AppendixA}, see \eqref{eq:FloquetHO}-\eqref{eq:FloquetFirstOrder}.

\subsubsection{Weak formulation and discretization}
To simplify the notation, we introduce the traces
\begin{subequations}
\begin{align}
\ensuremath{\mathbf{u}}_\mathrm{T} &\coloneqq \left(\ensuremath{\mathbf{n}}_{{\ensuremath{\Gamma}}} \times \ensuremath{\mathbf{u}}\vert_{{\ensuremath{\Gamma}}}  \right) \times \ensuremath{\mathbf{n}}_{{\ensuremath{\Gamma}}},\\
\ensuremath{\mathbf{u}}_\mathrm{t} &\coloneqq \ensuremath{\mathbf{n}}_{{\ensuremath{\Gamma}}} \times \ensuremath{\mathbf{u}}  \vert_{{\ensuremath{\Gamma}}},
\end{align}
\end{subequations}
where ${\ensuremath{\Gamma}}\coloneqq\partial\ensuremath{D}$ denotes the boundary of $\ensuremath{D}$ and $\ensuremath{\mathbf{n}}_{{\ensuremath{\Gamma}}}$ refers to its outer unit normal. Note that the trace operators are denoted by subscripts, for brevity of notation.

By building the inner product of \eqref{eq:MaxwellBVP_vol} with tests function $\ensuremath{\mathbf{E}}' \in V$, where $V$ is to be determined, and integration by parts we obtain 
\begin{align}
\left( \mu_\text r^{-1} \nabla \times \ensuremath{\mathbf{E}} , \nabla \times \ensuremath{\mathbf{E}}'\right)_\ensuremath{D}  - \omega^2 \mu_0 \left(\varepsilon \ensuremath{\mathbf{E}} , \ensuremath{\mathbf{E}}'\right)_\ensuremath{D} -j\omega \mu_0 \left( \ensuremath{\mathbf{H}}_\mathrm{t} , \ensuremath{\mathbf{E}}'_\mathrm{T} \right)_{\ensuremath{\Gamma}} =0. \label{eq:weakOhneSpaces}
\end{align}
The boundary integral can be further simplified, i.e. the contributions on $\ensuremath{{\Gamma_{x^+}}}$, $\ensuremath{{\Gamma_{x^-}}}$ and $\ensuremath{{\Gamma_{y^+}}}$, $\ensuremath{{\Gamma_{y^-}}}$ cancel each other due to the periodic phase-shift boundary conditions \eqref{eq:MaxwellBVP_P1}, \eqref{eq:MaxwellBVP_P2} of trial and test functions. We further eliminate the portion of the integral on $\ensuremath{{\Gamma_{z^-}}}$  by demanding that the test functions $\ensuremath{\mathbf{E}}'$ fulfill the \gls{pec} boundary condition \eqref{eq:MaxwellBVP_pec}. 

The appropriate function space $V$ for a weak formulation is a subspace of {\ensuremath{H\left(\mathrm{curl};\ensuremath{D}\right)}}, i.e. the (complex) vector function space of square-integrable functions with square-integrable curl. For more details on function spaces in the context of Maxwell's source problem, the reader is referred to \cite[Chapter 3]{monk2003}. To account for the boundary conditions in \eqref{eq:MaxwellBVP}, the function space is chosen as
\begin{align}
\begin{split}
V:=\{&\ensuremath{\mathbf{v}} \in {\ensuremath{H\left(\mathrm{curl};\ensuremath{D}\right)}}:~\ensuremath{\mathbf{v}}_\mathrm{T} \vert_\ensuremath{{\Gamma_{z^-}}} =0 \land  \  {\ensuremath{\mathbf{v}}_\mathrm{T}}\vert_{\ensuremath{{\Gamma_{x^+}}}} = -{\ensuremath{\mathbf{v}}_\mathrm{T}}\vert_{\ensuremath{{\Gamma_{x^-}}}} e^{j\psi_x} \ \\
&\land{\ensuremath{\mathbf{v}}_\mathrm{T}}\vert_{\ensuremath{{\Gamma_{y^+}}}} = -{\ensuremath{\mathbf{v}}_\mathrm{T}}\vert_{\ensuremath{{\Gamma_{y^-}}}} e^{j\psi_y} \land \ \ensuremath{\mathbf{v}}_\mathrm{T} \vert_{\ensuremath{{\Gamma_{z^+}}}} \in \left(L^2(\ensuremath{{\Gamma_{z^+}}})\right)^3\},\label{eq:V}
\end{split}
\end{align}
where the condition $\ensuremath{\mathbf{v}}_\mathrm{T} \vert_{\ensuremath{{\Gamma_{z^+}}}} \in \left(L^2(\ensuremath{{\Gamma_{z^+}}})\right)^3$ is required to obtain a well-defined boundary integral. Employing the Floquet absorbing boundary condition \eqref{eq:MaxwellBVP_Floquet} on $\ensuremath{{\Gamma_{z^+}}}$ yields the weak formulation: find $\ensuremath{\mathbf{E}} \in V$ s.t. 
\begin{align}
\begin{split}
\left( \mu_\text r^{-1} \nabla \times \ensuremath{\mathbf{E}} , \nabla \times \ensuremath{\mathbf{E}}'\right)_\ensuremath{D}  - \omega^2 \mu_0 \left(\varepsilon \ensuremath{\mathbf{E}} , \ensuremath{\mathbf{E}}'\right)_\ensuremath{D} +j\omega \ensuremath{\mu_0}\left(\boldsymbol{\mathcal G}(\ensuremath{\mathbf{E}}), \ensuremath{\mathbf{E}}'_\mathrm{T} \right)_\ensuremath{{\Gamma_{z^+}}} =j\omega\ensuremath{\mu_0} \left(\boldsymbol{\mathcal F}^\text{inc} , \ensuremath{\mathbf{E}}'_\mathrm{T} \right)_\ensuremath{{\Gamma_{z^+}}} \quad \forall \ensuremath{\mathbf{E}}' \in V\label{eq:WeakFormI},
\end{split}
\end{align}

To ensure a $\text{curl}$-conforming discretization of \eqref{eq:WeakFormI}, we approximate the electric field $\ensuremath{\mathbf{E}}$ numerically as
\begin{align}
\ensuremath{\mathbf{E}}_h(\ensuremath{\mathbf{x}}) = \sum_{j=1}^{\ensuremath{{N_h}}} c_j \ensuremath{\mathbf{N}}_j(\ensuremath{\mathbf{x}})
\label{eq:nedelec}
\end{align}
where $\ensuremath{\mathbf{N}}_j$ denotes N{\'e}d{\'e}lec basis functions of the first kind \cite{nedelec1980,monk2003} and 1st or 2nd order, defined on a tetrahedral mesh of the domain $\ensuremath{D}$. Further details on the discretization are given in \ref{sec:AppendixB}.

In practice, one is often interested in reflection and transmission coefficients, in addition to the field solution $\ensuremath{\mathbf{E}}$ itself. Therefore, we define the (complex-valued) scattering parameters as \mbox{(affine-)} linear functionals of $\ensuremath{\mathbf{E}}$
\begin{align}
S_{\alpha, mn} \coloneqq \left(\ensuremath{\mathbf{E}}_\mathrm{T}-\ensuremath{\ensuremath{\mathbf{E}}^\mathrm{inc}}_\mathrm{T},\ensuremath{\pi_\mathrm T}\left[\ensuremath{\mathbf{E}}_{\ensuremath{\alpha},mn}\right]\right)_\ensuremath{{\Gamma_{z^+}}}, \label{eq:Spara}
\end{align}
where $\alpha\in\{\text{TE},\text{TM}\}, m\in \mathbb Z, n\in \mathbb Z$ and $\ensuremath{\mathbf{E}}_{\ensuremath{\alpha},mn}$ are Floquet modes defined in \ref{sec:AppendixA}. The scattering parameters are considered as \glspl{qoi}, in the context of the present work. 

\subsection{Parametrized model}
\label{sec:parametrized_model}
In this subsection we specify the material distribution of the complex permittivity $\varepsilon$. 
In particular, we assume a linear material behaviour for $\varepsilon$ and $\mu$ inside $\ensuremath{D}$.  
Let the domain $\ensuremath{D}$ be composed of $M$ non-overlapping subdomains $D_m$, i.e. $\overline{\ensuremath{D}}=\bigcup_{m=1}^M \overline{D_m}$. 
We further assume that the dispersive permittivity $\varepsilon(\ensuremath{\mathbf{x}}, \omega)$ is spatially piecewise constant on each subdomain $D_m$
and depends smoothly on a given vector of $N$ parameters $\ensuremath{\mathbf{y}}\in \Xi \subset \mathbb R^N$
\begin{align}
\varepsilon(\ensuremath{\mathbf{x}}, \omega, \ensuremath{\mathbf{y}})=\sum_{m=1}^{M}\varepsilon_m(\omega, \ensuremath{\mathbf{y}}) \mathbbm{1}_m(\ensuremath{\mathbf{x}}, \ensuremath{\mathbf{y}}), \quad \mathrm{where} ~ \mathbbm{1}_m(\ensuremath{\mathbf{x}}, \ensuremath{\mathbf{y}})=\begin{cases}
1~,~ \ensuremath{\mathbf{x}}\in D_m(\ensuremath{\mathbf{y}}),\\
0~, ~\ensuremath{\mathbf{x}}\notin D_m(\ensuremath{\mathbf{y}}).\end{cases}\label{eq:permittivity}
\end{align} 
On the one hand, the parameter  vector $\ensuremath{\mathbf{y}}$ can be used to represent variations in the material parameters, e.g. different permittivities, refractive indices or extinction coefficients, by changing the coefficients $\varepsilon_m(\omega, \ensuremath{\mathbf{y}})$. 
On the other hand, it also represents geometric variations of the structure inside the unit cell, since the subdomains $D_m(\ensuremath{\mathbf{y}})$ for each material depend on $\ensuremath{\mathbf{y}}$ as well.

The parametrized weak formulation reads: find $\ensuremath{\mathbf{E}}(\ensuremath{\mathbf{y}})\in V$ s.t. 
\begin{align}
a_{\ensuremath{\mathbf{y}}}(\ensuremath{\mathbf{E}}(\ensuremath{\mathbf{y}}), \ensuremath{\mathbf{E}}') = l(\ensuremath{\mathbf{E}}') ~\forall \ensuremath{\mathbf{E}}'\in V,
\label{eq:WeakFormulationY}
\end{align}
where 
\begin{align}
a_{\ensuremath{\mathbf{y}}}(\ensuremath{\mathbf{E}}, \ensuremath{\mathbf{E}}') \coloneqq \left( \mu_\text r^{-1} \nabla \times \ensuremath{\mathbf{E}}(\ensuremath{\mathbf{y}}) , \nabla \times \ensuremath{\mathbf{E}}'\right)_\ensuremath{D} - \omega^2 \mu_0 \left(\varepsilon(\ensuremath{\mathbf{y}}) \ensuremath{\mathbf{E}}(\ensuremath{\mathbf{y}}), \ensuremath{\mathbf{E}}'\right)_\ensuremath{D}+j\omega \ensuremath{\mu_0}\left(\boldsymbol{\mathcal G}\bigl(\ensuremath{\mathbf{E}}(\ensuremath{\mathbf{y}})\bigr), \ensuremath{\mathbf{E}}'_\mathrm{T} \right)_\ensuremath{{\Gamma_{z^+}}}.\label{eq:bilinear_ho_Y}
\end{align}
The parameter-dependent scattering parameters are given as
\begin{align}
\begin{split}
S_{\ensuremath{\alpha},mn}(\ensuremath{\mathbf{y}}) = \left(\ensuremath{\mathbf{E}}_\mathrm{T}(\ensuremath{\mathbf{y}})-\ensuremath{\ensuremath{\mathbf{E}}^\mathrm{inc}}_\mathrm{T},\ensuremath{\pi_\mathrm T}\left[\ensuremath{\mathbf{E}}_{\ensuremath{\alpha},mn}\right]\right)_\ensuremath{{\Gamma_{z^+}}},\quad \text{~where~}\ensuremath{\alpha} \in\{\text{TE, TM\}}. \label{eq:SParaY}
\end{split}
\end{align}

\begin{remark}
Relating the model problem of scattering in periodic media to the \gls{uq} methodology of the previous section, the linear functional $J_{\mathbf y}(\cdot)$ is given by
\begin{align*}
S_{\ensuremath{\alpha},mn}(\ensuremath{\mathbf{y}}) = \underbrace{\left(\ensuremath{\mathbf{E}}_\mathrm{T}(\ensuremath{\mathbf{y}}),\ensuremath{\pi_\mathrm T}\left[\ensuremath{\mathbf{E}}_{\ensuremath{\alpha},mn}\right]\right)_\ensuremath{{\Gamma_{z^+}}}}_{=J_{\mathbf y}(\ensuremath{\mathbf{E}})}-\left(\ensuremath{\ensuremath{\mathbf{E}}^\mathrm{inc}}_\mathrm{T},\ensuremath{\pi_\mathrm T}\left[\ensuremath{\mathbf{E}}_{\ensuremath{\alpha},mn}\right]\right)_\ensuremath{{\Gamma_{z^+}}},
\end{align*}where $\ensuremath{\alpha}\in\{\mathrm{TE}, \mathrm{TM}\}$.  
The strong formulation of the adjoint problem \eqref{eq:dual_problem} reads 
\begin{subequations}
\begin{alignat}{2}
\nabla \times \left(\frac 1 {\mu_r^*} \nabla \times \ensuremath{\mathbf{z}}\right)- \omega^2 \mu_0 \varepsilon^* \ensuremath{\mathbf{z}}&= 0 & &\mathrm{in}~\ensuremath{D},\\
\ensuremath{\mathbf{z}}_\mathrm{T}\vert_{\ensuremath{{\Gamma_{x^+}}}} e^{-j\psi_x}&= \ensuremath{\mathbf{z}}_\mathrm{T}\vert_{\ensuremath{{\Gamma_{x^-}}}}   & &\mathrm{on}~{\ensuremath{{\Gamma_{x^+}}}\cup \ensuremath{{\Gamma_{x^-}}}}, \\
\ensuremath{\mathbf{z}}_\mathrm{T}\vert_{\ensuremath{{\Gamma_{y^+}}}} e^{-j\psi_y}&= \ensuremath{\mathbf{z}}_\mathrm{T}\vert_{\ensuremath{{\Gamma_{y^-}}}}  & &\mathrm{on}~{\ensuremath{{\Gamma_{y^+}}}\cup \ensuremath{{\Gamma_{y^-}}}},\\
\ensuremath{\mathbf{z}}_\mathrm{t} &= 0 & &\mathrm{on~}\ensuremath{{\Gamma_{z^-}}},\\
\ensuremath{\ensuremath{\mathbf{e}}_z}\times\left(\frac{j}{\omega \ensuremath{\mu_0}} \nabla \times \ensuremath{\mathbf{z}}\right)
+ \boldsymbol{ \overline{\mathcal G}} &= \boldsymbol{\overline{\mathcal F}}
& &\mathrm{on}~\ensuremath{{\Gamma_{z^+}}},
\end{alignat}\label{eq:dual_strong}\end{subequations}
where $\boldsymbol{ \overline{\mathcal G}}$ and $\boldsymbol{\overline{\mathcal F}}$ are defined in \ref{sec:AppendixA}. 

Introducing $\mathbf A_{\mathrm{dof}}$ as system matrix arising from the discretization of \eqref{eq:WeakFormI}, the discrete primal problem is given by 
\begin{equation}
\mathbf A_{\mathrm{dof}} \mathbf c_{\mathrm{dof}} = \mathbf f_{\mathrm{dof}},
\end{equation}
as derived in \ref{sec:AppendixB}. Discretization of the adjoint problem \eqref{eq:dual_problem} yields the discrete matrix equation
\begin{align}
\mathbf{A}^\mathrm H_\ensuremath{\mathrm{dof}} \ensuremath{\mathbf{z}}_\ensuremath{\mathrm{dof}}= {\ensuremath{\mathbf{J}}_\ensuremath{\mathrm{dof}}},  \label{eq:DiscreteSystemDual}
\end{align}
and the discrete version of the error indicator \eqref{eq:errorEstContinuous} reads
\begin{align}
\tilde \eta_h(\ensuremath{\mathbf{y}}) = \tilde{\ensuremath{\mathbf{z}}}_\ensuremath{\mathrm{dof}}^\mathrm H(\ensuremath{\mathbf{y}})\ensuremath{\mathbf{f}}_\ensuremath{\mathrm{dof}}(\ensuremath{\mathbf{y}}) - \tilde{\ensuremath{\mathbf{z}}}_\ensuremath{\mathrm{dof}}^\mathrm H(\ensuremath{\mathbf{y}}) \mathbf{A}_\ensuremath{\mathrm{dof}}(\ensuremath{\mathbf{y}}) \tilde{\ensuremath{\mathbf{c}}}_\ensuremath{\mathrm{dof}}(\ensuremath{\mathbf{y}}). \label{eq:errorEstDiscrete}
\end{align}

Note that the dual solution can be obtained with negligible cost in many cases, e.g. if the primal problem is solved with a sparse LU decomposition $\mathbf{A}_\ensuremath{\mathrm{dof}} = \mathbf{L}\mathbf{U}$, for the respective dual problem we obtain \begin{equation}\mathbf{A^\mathrm{H}_\ensuremath{\mathrm{dof}}}=\left(\mathbf{L}\mathbf{U}\right)^\mathrm{H}=\mathbf{U}^\mathrm{H}\mathbf{L}^\mathrm{H}.\end{equation}

\end{remark}

\begin{remark}
To prove that problem \eqref{eq:WeakFormulationY} is well-posed and analytic with respect to the model parameters, which are the main working assumptions of the paper, requires special care due to the presence of the general boundary operator $\mathcal{G}$. Here we only refer to \cite[Chapter 4]{monk2003} and \cite[Section 5]{Hiptmair_2002aa} for a numerical analysis of well-posedness in the deterministic setting with simpler boundary conditions. Also, recently, shape holomorphy for Maxwell's source problem was established in \cite{aylwin2019domain}, considering bi-Lipschitz shape transformations and holomorphic material parameters, which are bounded away from zero. However, the analysis was, again, carried out with homogeneous Dirichlet boundary conditions. 
\end{remark}
\section{Application}
\label{sec:aplication}

We apply the enhanced surrogate modeling presented in Section \ref{sec:uq} to a non-trivial benchmark application from nanoplasmonics, namely an optical grating coupler \cite{preiner2008, cstTutorial}.
We report some details on modeling uncertainties in material and geometric input data. We also describe how parametric variations are realized numerically and finally quantify uncertainties in the computational model.

For general periodic structures, we must distinguish between two types of uncertainties. 
In this work we focus on \textit{global} uncertainties, 
i.e. we assume that all unit cells are identically affected, modeling a systematic offset in the fabrication process. 
We do not address \textit{local} uncertainties leading to a violation of the periodicity and different unit cells. 
Readers interested in the latter case are referred to \cite{schmitt2019optimization} for a relevant study.

\subsection{Numerical model}
\begin{figure}
\begin{subfigure}[b]{.45\textwidth}
\resizebox{.95\textwidth}{!}{\includegraphics{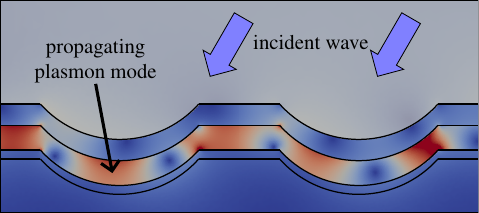}} \vspace{.2em}	\caption{~}
\label{fig:mimPlasmonMode}
\end{subfigure}
\begin{subfigure}[b]{.54\textwidth}
\begin{flushright}
\resizebox{.95\textwidth}{!}{\includegraphics{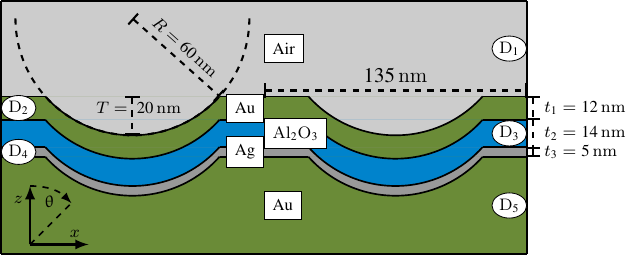}}\vspace{-\baselineskip}
\end{flushright}
\caption{~}
\label{fig:design_coupler}
\end{subfigure}
\caption{An optical grating coupler \cite{preiner2008} couples power from an incident plane wave in free space directly into a \gls{mim} plasmon mode, propagating in horizontal direction. This is illustrated in Fig. ~\ref{fig:mimPlasmonMode}, where the coloring indicates the magnitude of the electric field at a specific (arbitrary chosen) point in time. Fig.~\ref{fig:design_coupler} shows the design of the considered grating coupler. Note that the structure is periodically extended in horizontal direction, here, only two unit cells are shown.}
\end{figure}

The considered grating \cite{preiner2008} couples power from an incident \gls{tm} polarized plane wave, such that
\begin{align*}
\ensuremath{\pi_\mathrm T}\left[\ensuremath{\ensuremath{\mathbf{E}}^\mathrm{inc}}\right] = \ensuremath{\pi_\mathrm T}\left[ \ensuremath{\mathbf{E}}_{\text{TM},00}\right], \quad~ \text{at~} \ensuremath{{\Gamma_{z^+}}},
\end{align*} with propagation direction  $\ensuremath{\theta^\mathrm{inc}}=\num{53}^\circ,~\ensuremath{\phi^\mathrm{inc}}=0^\circ$, directly into a \gls{mim} plasmon mode, which is illustrated in Fig.~\ref{fig:mimPlasmonMode}.

The structure's design, shown in Fig.~\ref{fig:design_coupler}, is assumed to be periodic in the $x$ direction and infinitely extended in the $y$ direction. 
The reflection coefficients \eqref{eq:Spara} at the upper boundary $\ensuremath{{\Gamma_{z^+}}}$ correspond to the coupling efficiency of the structure, such that larger reflection coefficients indicate a lower coupling efficiency. Therefore, the scattering parameter $S:=S_{\text{TM},00}$ is considered as the \gls{qoi} in the following. 
Note that we focus on the fundamental reflection coefficient $S_{\text{TM},00}$ because, for this particular model, all other scattering parameters have negligible amplitudes.

\begin{figure*}
\centering
\includegraphics{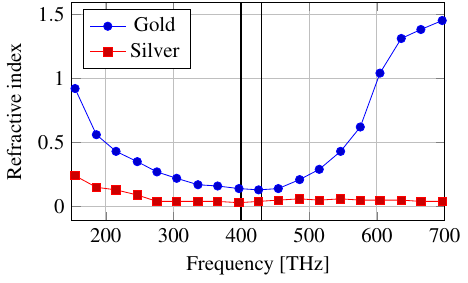}\hfill\includegraphics{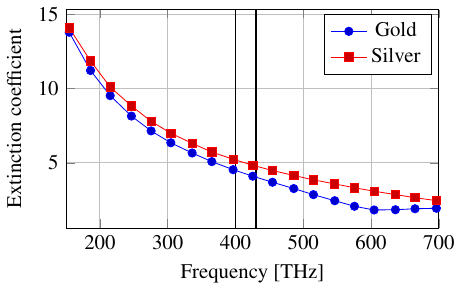}
\caption{Dispersive optical constants of gold and silver \cite{johnson1972}. The black vertical lines define the considered frequency range.}
\label{fig:johnsonData}
\end{figure*}

We model the material properties based on measurement data for noble metals provided by Johnson and Christy in \cite{johnson1972} and presented in Table~\ref{table:MaterialData}. 
We focus on the frequency range $f_{\min}=\SI{400}{THz}$ to $f_{\max}=\SI{430}{THz}$, see Fig.~\ref{fig:johnsonData}.
The data is experimentally determined by reflectivity studies and is therefore given in terms of the refractive indices $n$ and the extinction coefficients $\kappa$ for gold and silver, respectively. 
From those, one obtains the complex permittivity as in \cite[Chapter 1.1]{maier2007}, i.e. 
\begin{equation} 
\varepsilon = \left(n^2 - \kappa^2 - j(2n\kappa)\right)\ensuremath{\varepsilon_0}.
\end{equation}
We interpolate the given material $\beta^\alpha_i$ data as
\begin{align}
\beta^\alpha(\omega) &= \sum_{i=0}^2 \beta_i^\alpha l_i(\omega),
\end{align}
where $\alpha \in \{ \text{Au},\text{Ag}\}$, $\beta \in \{n,\kappa \}$. Also,
\begin{equation}l_i(\omega) = \prod_{j=0, j\neq i}^{2} \frac {\omega-\omega_j}{\omega_i-\omega_j}, ~~\omega_i=2\pi f_i,\end{equation}
are 2nd order Lagrange polynomials and $f_i, n_i^\text{Au}, \kappa_i^\text{Au}, n_i^\text{Ag}$, $\kappa_i^\text{Ag}$, $i=0, 1,2$ are given.

\begin{table*}
\caption{Material data, taken from \cite{johnson1972}.}
\label{table:MaterialData}
\centering
\begin{tabular}{|c|c|c|c|c|c|c|}
\hline
Index & Energy& Frequency & Refractive & Extinction  & Refractive & Extinction \\
$i$&[\SI{}{eV}]& $f_i$ [\SI{}{THz}] & index $n^\text{Au}_i$& coefficient $\kappa^\text{Au}_i$ & index $n^\text{Ag}_i$ & coefficient      $\kappa^\text{Ag}_i$\\\hline
0&1.64 & 396.55 & $0.14\pm 0.02$ & $4.542\pm0.015$ & $0.03 \pm 0.02$ & $5.242 \pm 0.015$ \\\hline
1&1.76 & 425.57 & $0.13\pm 0.02$ & $4.103\pm0.010$ & $0.04 \pm 0.02$ & $4.838 \pm 0.010$ \\\hline
2&1.88 & 454.58 & $0.14\pm 0.02$ & $3.697\pm0.007$ & $0.05 \pm 0.02$ & $4.483 \pm 0.007$ \\\hline
\end{tabular}
\end{table*}

We proceed with the description of the deterministic numerical model, as well as its parametrization. 
The periodic mesh for the nominal design is created using \textsc{Gmsh} \cite{geuzaine2009}. Since for this particular structure only the fundamental Floquet modes propagate and all higher order modes are attenuated to a negligible amplitude at $\ensuremath{{\Gamma_{z^+}}}$, we can use the first order Floquet boundary condition \eqref{eq:FloquetFirstOrder}. 
We use \mbox{\textsc{FEniCS}} \cite{alnaes2015} as \gls{fe} library to assemble the \gls{fe} matrix $\mathbf A$ and \gls{rhs} $\ensuremath{\mathbf{f}}$ (see \eqref{eq:DiscreteSystem}), as well as the linear functional $\ensuremath{\mathbf{J}}_\ensuremath{\mathrm{dof}}$ used for the numerical approximation of the scattering parameter
\begin{align*}
S_{\text{TM},00} =  \left(\ensuremath{\mathbf{E}}_\mathrm{T}- \ensuremath{\ensuremath{\mathbf{E}}^\mathrm{inc}}_\mathrm{T}, \ensuremath{\pi_\mathrm T}\left[\ensuremath{\mathbf{E}}_{\text{TM},00}\right]\right)_\ensuremath{{\Gamma_{z^+}}} \approx \ensuremath{\mathbf{J}}_\ensuremath{\mathrm{dof}}^{\mathrm{H}}\ensuremath{\mathbf{c}}_\ensuremath{\mathrm{dof}}-1.
\end{align*} 
Since \textsc{FEniCS 2017.2.0} is not able to deal with complex numbers, we assemble the real and the imaginary parts of the matrix and the vectors separately. 
We then use \textsc{numpy} and \textsc{scipy} to impose the quasi-periodic boundary conditions \eqref{eq:phaseshiftBCx}, \eqref{eq:phaseshiftBCy} and solve the resulting linear system \eqref{eq:DiscreteSystem} with a sparse LU decomposition. 
Using 2nd order N{\'e}d{\'e}lec elements of the 1st kind, we end up with \num{56200} \gls{dof}s and achieve an accuracy of $\approx \num{e-3}$ in the scattering parameter. The reference solutions for different frequency sample points are computed with a commercial software \cite{cst} employing an adaptively refined mesh of higher order curved elements.
Since a sparse LU decomposition is used to solve the resulting linear system, the adjoint solution $\ensuremath{\mathbf{z}}_\ensuremath{\mathrm{dof}}$ is obtained with negligible costs.

\begin{figure*}
\centering
\begin{subfigure}[b]{0.3\textwidth}
\includegraphics{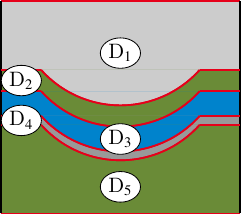}
\caption{Subdomains}
\end{subfigure}
\begin{subfigure}[b]{0.3\textwidth}
\includegraphics{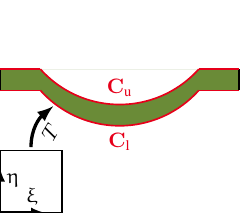}
\caption{Mapping}
\end{subfigure}
\begin{subfigure}[b]{0.19\textwidth}
\centering
\includegraphics[width=1\textwidth, trim = 0em 10em 0px 0px, clip]{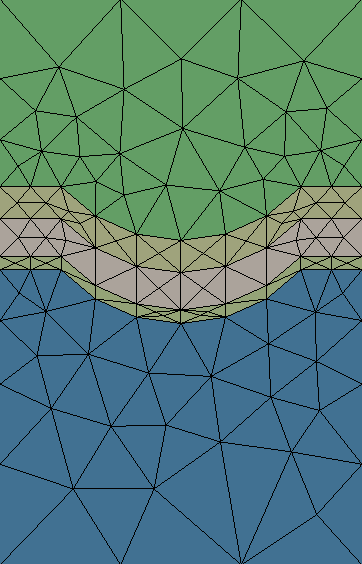}
\caption{Initial}
\end{subfigure}
\begin{subfigure}[b]{0.19\textwidth}
\centering
\includegraphics[width=1\textwidth, trim = 0em 10em 0px 0px, clip]{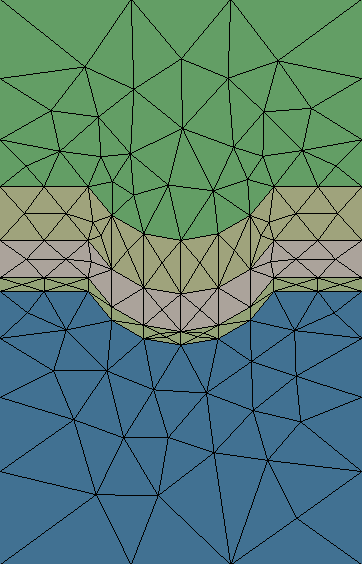}
\caption{Deformed}
\end{subfigure}
\caption{a) design elements. b) mapping from unit square. c) initial mesh (coarse for illustration) for nominal design $\ensuremath{\mathbf{y}}^\text{nominal}$. d) deformed mesh for $R=\SI{40}{nm},~t_1=\SI{20}{nm}$.}
\label{fig:design_elements}
\end{figure*} 
To incorporate changes in the geometry parameters without the need to re-mesh, a design element approach is applied \cite{braibant1984}. 
We describe all (material) interfaces, illustrated in Fig.~\ref{fig:design_elements} in red color, using \gls{nurbs} \cite{piegl1997}. Each \gls{nurbs} curve
\begin{align}
\ensuremath{\ensuremath{\mathbf{C}}}_i(\xi;\ensuremath{\mathbf{y}})= \sum_{j=0}^{n} R_j(\xi) \ensuremath{\mathbf{P}}_{j,i}(\ensuremath{\mathbf{y}}), ~~\xi \in [0,1]
\end{align}
is a superposition of rational basis functions $R_j(\xi)$ weighted by control points $\ensuremath{\mathbf{P}}_j$. We then define mappings
\begin{align}
\ensuremath{\mathbf{T}}_m(\xi, \eta; \ensuremath{\mathbf{y}}) = \eta \ensuremath{\ensuremath{\mathbf{C}}}_{m,\text{u}}(\xi; \ensuremath{\mathbf{y}}) + (1-\eta)\ensuremath{\ensuremath{\mathbf{C}}}_{m,\text{l}}(\xi; \ensuremath{\mathbf{y}}),
m = 1,\ldots,M, \label{eq:mapping}
\end{align}
from the unit square $[0\le \xi \le 1]\times [0\le \eta \le 1]$ to each design element $D_i(\ensuremath{\mathbf{y}})$ (see Fig.~\ref{fig:design_elements}). Thereby, the subscripts $\text{u}$ and $\text{l}$ refer to the upper and lower \gls{nurbs} curve of the design element, respectively. Given the initial mesh, for each mesh node the respective coordinates on the unit square are found by solving a non-linear root-finding problem. 
We can then deform the mesh by moving the mesh nodes to the new coordinates obtained by evaluating the mapping \eqref{eq:mapping} for different geometry parameters $\ensuremath{\mathbf{y}}$.

\renewcommand{\arraystretch}{1.2}
\begin{table}
\caption{Uncertain geometrical parameters.}
\label{tab:SingleFreqParameters}
\centering
\begin{tabular}{|c|c|c|}
\hline
\textbf{Parameter} & \textbf{Nominal value} & \textbf{Variation} \\
\hline Grating radius $R$ & \SI{60}{nm} & $\pm$\SI{1.5}{nm}\\
\hline Gold layer thickness $t_1$ & \SI{12}{nm} & $\pm$ \SI{1.5}{nm} \\
\hline Alumina layer thickness $t_2$ & \SI{14}{nm} & $\pm$ \SI{1.5}{nm}\\
\hline Silver layer thickness $t_3$ & \SI{5}{nm} & $\pm$ \SI{1.5}{nm} \\
\hline Grating depth $T$ & \SI{20}{nm} & $\pm$ \SI{1.5}{nm} \\\hline 
\end{tabular}
\end{table}

We consider a fixed frequency $\omega = 2\pi \left(\SI{414}{THz}\right)$ and $N=17$ random input parameters $\ensuremath{\mathbf{Y}}$. These are the 5 geometrical parameters presented in Table~\ref{tab:SingleFreqParameters} and the 12 material parameters given in Table~\ref{table:MaterialData}. As introduced in Section~\ref{sec:maxwell}, both the uncertain geometry and the uncertain material coefficients are modeled by an uncertain complex permittivity $\varepsilon(\ensuremath{\mathbf{x}}, \ensuremath{\mathbf{y}})$, see \eqref{eq:permittivity}.

We assume that the \gls{rv}s $Y_n$, $n=1,2,\dots,N=17$, are independent and distributed in the ranges defined by their nominal values and variations. The variations of the material parameters are chosen according to the error estimate provided by Johnson and Christy  \textit{based on the instrumental accuracy of the reflection and transmission measurements} \cite{johnson1972}. Since no further information on the distributions of those measurement uncertainties are specified, the given error estimate is assumed to correspond to a $2\sigma$ interval. For the geometrical parameters only small variations in the range of $\pm \SI{1.5}{nm}$ are considered (with a $2\sigma$ interval of $\pm\SI{1}{nm}$).

We opt for beta distributions, which have bounded support and can approximate normal distributions for suitable choices of their shape parameters \cite[Appendix B]{xiu2010}. The shape parameters are chosen based on the results of a series of Kolmogorov-Smirnov fitting tests \cite{lopes2011}. 

\subsubsection{Numerical Results}
\begin{figure*}
\centering
\begin{subfigure}[b]{.48\textwidth}
\includegraphics{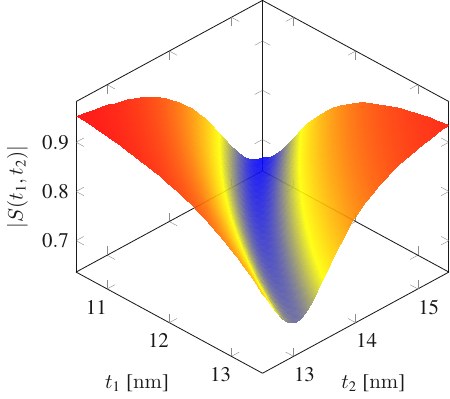}
\vspace{.7em}
\caption{}
\label{fig:scattering}
\end{subfigure}\hfill
\begin{subfigure}[b]{.48\textwidth}
\includegraphics{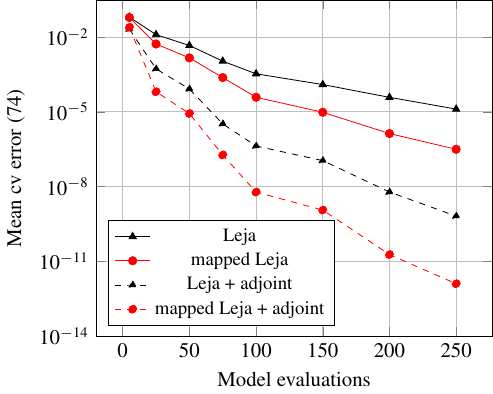}
\caption{}
\label{fig:conv2d}
\end{subfigure}
\caption{Considering a two dimensional parameter space and a fixed frequency of $\SI{414}{THz}$. a) Reflection coefficient $|S|$. b) Improved convergence of (mapped) Leja approximations by employing an adjoint-error indicator.}
\end{figure*}

To illustrate the benefits of using the adjoint error indicator presented in Sec.~\ref{sec:adjointAdapt}, we consider here only the thickness of the upper gold layer $t_1$ and the thickness of the dielectric layer $t_2$ as input parameters. We have observed numerically, that the \gls{qoi} is particularly sensitive with respect to these parameters, with slow associated univariate convergence rates. Fig.~\ref{fig:scattering} shows the S-parameter w.r.t. to small variations of these two parameters. 
We construct (mapped) adaptive Leja approximations using Algorithm~\ref{algo:gensmolyak} and the adjoint-based Algorithm~\ref{algo:gensmolyak2}.
The accuracy of the surrogate models is measured using a cross-validation set of $N^\text{cv}=1000$ parameter realizations $S^{(i)}:=S(\ensuremath{\mathbf{y}}^{(i)}), ~i=1,\ldots,N^\text{cv}$, drawn according to the underlying PDF, which is used to compute a discrete approximation of the $L^1_\varrho$ error
\begin{align}
\mathbb{E}[|S-\tilde{S}|] \approx 
\frac{1}{N^\text{cv}}\sum_{i=1}^{N^\text{cv}} |S^{(i)}-\tilde{S}^{(i)}|. \label{eq:mean_error}
\end{align}
The error \eqref{eq:mean_error} is computed for the mapped and adjoint-based  approximation, as well as for non-mapped and/or non-adjoint-based variants, for increasing numbers of model evaluations.  
The corresponding results are shown in Fig.~\ref{fig:conv2d}.
The plot numerically confirms \eqref{eq:adjoint_conv} and shows the doubled convergence order of the adjoint-error indicator. 
Additionally, it can already be observed that employing the conformal sausage map $\ensuremath{g_\mathrm{S}}(\cdot;9)$, defined in \eqref{eq:sausage_map}, yields to a significant improvement of both convergence orders in the considered setting.

Next, we consider all $N=17$ input parameters and construct different polynomial approximations. As a reference, we compute two non-adaptive approximations, based on \gls{gpc} \cite{xiu2002} and on isotropic Smolyak sparse-grid interpolation \cite{back2011}. These are compared to the proposed enhanced surrogate modeling, i.e. (mapped) Leja adaptive approximations, using both Algorithm \ref{algo:gensmolyak} and the adjoint-based Algorithm \ref{algo:gensmolyak2} for the latter.
\textsc{Chaospy} \cite{feinberg2015} is used for the \gls{gpc} case, the \textsc{sparse-grid-matlab-kit} \cite{back2011} is employed for the Smolyak sparse-grid interpolation, while an in-house code was developed for both Leja adaptive algorithms\footnote{\url{https://github.com/dlouk/DALI3}}. 
We compare the resulting surrogate models with respect to accuracy and computational costs.

The computational costs refer to the number of model evaluations needed for the construction of the approximation.
While straightforward for the \gls{gpc}, Smolyak sparse-grid and the Leja adaptive Algorithm \ref{algo:gensmolyak}, the estimation of costs is more involved in the case of the adjoint-based Algorithm \ref{algo:gensmolyak2}. 
First, in order to evaluate the duality-based error indicator \eqref{eq:errorEstDiscrete} at a candidate point, it is sufficient to evaluate a residual of \eqref{eq:WeakFormulationY}. 
Therefore, we distinguish between residual evaluations and solver calls, where in most cases the costs to evaluate the residuals are 
almost negligible compared to the solver costs, i.e. assembly and sparse LU decomposition of the system matrices $\mathbf{A}_\ensuremath{\mathrm{dof}}(\ensuremath{\mathbf{y}})$.
Second, the additional costs for computing the dual solution $\ensuremath{\mathbf{z}}$ by forward and backward substitution can also be neglected in most cases, since the primal problem is solved with a sparse LU decomposition.

As before, the accuracy of the surrogate models is measured using a cross-validation set of $N^\text{cv}=1000$ parameter realizations $S^{(i)}:=S(\ensuremath{\mathbf{y}}^{(i)}), ~i=1,\ldots,N^\text{cv}$, drawn according to the underlying PDF. In addition to \eqref{eq:mean_error}, we also consider the maximum error over all sample points
\begin{align}
\max_{i=1,\ldots, N^{\text{cv}}} |S^{(i)}-\tilde{S}^{(i)}|. \label{eq:max_error}
\end{align}
\begin{table*}
\caption{Accuracy and computational cost of different polynomial approximations for 17 input \glspl{rv}. \#LU refers to the dominating costs for the assembly and sparse LU decomposition of the system matrices. \#FB and \#Res denote the number of forward-backward substitutions and residual evaluations, respectively.}
\label{tab:17DgPC}
\centering
\begin{tabular}{|c|c|c|c|c|c|c|}
\hline
&  \textbf{\#LU} & \#FB & \#Res & Max. Error \eqref{eq:max_error} & Mean Error \eqref{eq:mean_error}\\\hline\hline
Total-degree gPC (without maps) & \textbf{\textit{613}} & \textit{613} & 0 &\num{7.61e-1}  &\num{1.92e-1}  \\\hline
Level 2 Smolyak sparse-grid (without maps) & \textbf{\textit{613}}& \textit{613} & 0 & \num{1.73e-1} & \num{4.45e-2} \\\hline \hline
Ad. Leja (without adjoints/maps)&  \textbf{\textit{613}} &  \textit{613} &0 &\num{8.56e-2} &\num{5.53e-3} \\\hline
Ad. iso-mapped Leja (without adjoints)&  \textbf{\textit{613}} &  \textit{613} &0 &\num{3.59e-2} &\num{4.10e-3} \\\hline
Ad. aniso-mapped Leja (without adjoints)&  \textbf{\textit{613}} &  \textit{613} &0 &\num{3.85e-2} &\num{4.15e-3} \\\hline\hline 
Ad. Leja (with adjoints; without maps)  & \textbf{558} & 1116 & \textit{613}& \num{8.46e-2}  & \num{5.49e-3}  \\ \hline
Ad. iso-mapped Leja (with adjoints)  & \textbf{563} &1126& \textit{613}& \num{3.57e-2}  & \num{4.09e-3}  \\ \hline
Ad. aniso-mapped Leja (with adjoints)  & \textbf{563} &1126& \textit{613}& \num{3.82e-2}  & \num{4.15e-3}  \\ \hline
\hline 
\makecell{Ad. aniso-mapped Leja \\(with adjoints and error correction)} & \textbf{3000} &6000& 30000& \num{1.25e-3}  & \num{1.20e-4} \\\hline 
\end{tabular}
\end{table*}

All accuracy and cost results are presented in Table~\ref{tab:17DgPC}.
First, a \gls{gpc} approximation with a 2nd order total-degree polynomial basis, i.e. \num{171} Jacobi polynomials, is constructed.
The polynomial coefficients are computed with a sparse 2nd order Gauss quadrature formula, resulting in \num{613} quadrature nodes, accordingly, model evaluations.
Second, we employ interpolation on an isotropic Smolyak sparse-grid of level 2 based on Clenshaw Curtis nodes, which requires \num{613} model evaluations as well.
Accordingly, we set a budget $B=613$ for the classical, i.e. without adjoints or conformal maps, Leja adaptive Algorithm \ref{algo:gensmolyak}, such that its costs are identical to the \gls{gpc} and the Smolyak sparse-grid interpolation.
As can be seen in Table~\ref{tab:17DgPC}, the Leja adaptive approximation is about one order of magnitude more accurate than the \gls{gpc} and also significantly better than the isotropic sparse-grid interpolation.

Next, we use again Algorithm~\ref{algo:gensmolyak} but employ conformal maps. In particular, we refer with \textit{iso-mapped} to applying the conformal sausage map $\ensuremath{g_\mathrm{S}}(\cdot;9)$ for all parameters while \textit{aniso-mapped} refers to the application of the conformal map only to the parameters $t_1, t_2, T$ (parameters with a particularly slow univariate convergence rate). It can be observed that both approaches yield a similar improvement in terms of accuracy, without (relevant) extra computational cost. 

For the adjoint-based Algorithm \ref{algo:gensmolyak2}, we then compute  approximations using again \num{613} (mapped) polynomials, resulting in errors almost identical to the non-adjoint case.
However, since the costs can be predominantly attributed to the $\approx \num{560}$ solver calls, the costs are reduced. 
Note that greater (relative) gains can be observed in different settings, e.g. when a smaller computational budget is used or less parameter anisotropy is present in the considered model. In particular, as a numerical test case, we  increased material uncertainties by one third and reduced geometric variations to the range of $\pm \SI{0.25}{nm}$. In that case the respective computational cost was reduced by more than $50\%$.

\begin{figure*}
\centering
\includegraphics{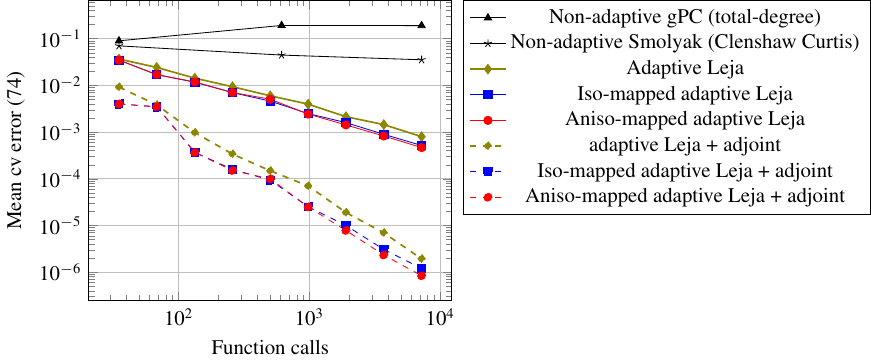}
\caption{Convergence study for the single frequency setting. The adaptive schemes clearly outperform the isotropic approach. Note that the dashed lines correspond to the error of \textit{improved} surrogate models which require the evaluation of FE residuals at each cross-validation point.}
\label{fig:sf_conv}
\end{figure*}

The convergence of the mean error \eqref{eq:mean_error} w.r.t. function calls (corresponding to the number of LU decompositions) of the investigated spectral methods is additionally shown in Fig.~\ref{fig:sf_conv}. Isotropic gPC does not seem to show a proper convergence. However, it should be noted that we were only able to compute approximations up to order 3 due to the larger number of parameters and hence, the error decay is probably pre-asymptotic. Additional convergence results supporting this hypothesis are reported in \ref{sec:AppendixC}. It can be concluded that the non-adaptive gPC reference solution achieves a very poor accuracy with the given computational budget. All considered dimension-adaptive schemes greatly outperform the isotropic approaches. In accordance with the results in Table~\ref{tab:17DgPC}, Fig.~\ref{fig:sf_conv} illustrates that the application of the conformal map leads to improvements compared to the classical Leja algorithm while, in this setting, there is a negligible difference between the \textit{iso-mapped} and the \textit{aniso-mapped} approach. It should be noted that, in contrast to Table~\ref{tab:17DgPC}, the dashed lines in Fig.~\ref{fig:sf_conv} correspond to the error of the respective adjoint-based error indicator \eqref{eq:errorEstDiscrete}. Therefore, they do not correspond to (mapped) polynomial surrogate models but require the evaluation of a residual of \eqref{eq:WeakFormulationY} at each cross-validation sample point $\mathbf{y}^{(i)},\,i=1,\ldots, N^\text{cv}$.

Finally, as shown in the last row of Table~\ref{tab:17DgPC}, we compute a very accurate surrogate model, by using the adjoint-based Algorithm \ref{algo:gensmolyak2} with a computational budget of \num{3000} LU decomposition. The adjoint-based approximation is then refined by employing \eqref{eq:approx_corrected} until $30000$ polynomials are used, further reducing the error by more than one order of magnitude. It shall be highlighted that the adjoint-based approach results in tremendous computational savings compared to the classical Leja Algorithm  \ref{algo:gensmolyak} since 27000 full model evaluations could be avoided in this particular setting.

As often pointed out in the literature, see e.g. \cite{jakeman2015}, it is inefficient to reduce the stochastic error below the discretization error. Therefore, the stochastic approximation is not further refined and the most accurate surrogate model (Table \ref{tab:17DgPC}, last row) is in the following used to compute statistical measures of the absolute value of the scattering parameter $|S|$. 

\subsubsection{Post-processing the surrogate model}
\label{subsubsec:postprocessing}
Since the (mapped) polynomial surrogate model $\tilde S(\ensuremath{\mathbf{y}})$ can be evaluated inexpensively, we employ a Monte Carlo-based approach
by evaluating the surrogate model on a large number of $N^\text{MC}$ parameter sample points, drawn from the joint \gls{pdf} $\varrho\left(\ensuremath{\mathbf{y}}\right)$.
We then use the sample evaluations to estimate statistical moments of $|S|$, its \gls{pdf}, failure probabilities based on specific design criteria, and its sensitivity with respect to the input parameters.

The expected value $\mathbb{E}[|S|]$ and the variance $\mathbb{V}[|S|]$ are estimated as \begin{subequations}
\begin{align}
\mathbb{E}[|S|] &= \int_{\Xi}|S(\ensuremath{\mathbf{y}})| \varrho\left(\ensuremath{\mathbf{y}}\right) \mathop{}\!\mathrm{d} \ensuremath{\mathbf{y}} \approx  \frac{1}{N^\text{MC}} \sum_{i=1}^{N^\text{MC}} |\tilde S^{(i)}|=:E^{\text{MC}}[|\tilde S|], \\
\mathbb{V}[|S|] &= {\int_{\Xi}\left(|S(\ensuremath{\mathbf{y}})|-\mathbb{E}[|S|]\right)^2 \varrho\left(\ensuremath{\mathbf{y}}\right) \mathop{}\!\mathrm{d} \ensuremath{\mathbf{y}}} \approx \frac{1}{N^\text{MC}-1} \sum_{i=1}^{N^\text{MC}} \left(|\tilde S^{(i)}|-E^{\text{MC}}[|\tilde S|]\right)^2.
\end{align}
\end{subequations}
We estimate the failure probability $\mathcal{F} = P(|S|\ge 1-\alpha)$ as
\begin{align}
\label{eq:MC_surrogate}
\mathcal F &= P(|S|\ge 1 - \alpha) =
\int_\Xi \mathcal{I}_\mathcal F\left(S(\ensuremath{\mathbf{y}})\right) \varrho\left(\ensuremath{\mathbf{y}}\right) ~\mathop{}\!\mathrm{d} \ensuremath{\mathbf{y}} \approx \frac{1}{N^\text{MC}} \sum_{i=1}^{N^\text{MC}} \mathcal{I}_\mathcal F\left(\tilde S^{(i)}\right), 
\end{align}where $\varrho_S$ denotes the \gls{pdf} of $|S|$ and $\mathcal I_\mathcal F$ denotes the indicator function
\begin{align}
\mathcal{I}_\mathcal F(S) = 	
\begin{cases}
1, ~~ |S|\in [1 - \alpha, 1],\\
0, ~~ |S|\in [0, 1 - \alpha).
\end{cases}
\end{align}
Monte Carlo sampling in combination with surrogate modeling is used for simplicity here. However, it should be noted that equality in \eqref{eq:MC_surrogate} for $N^{\text{cv}} \rightarrow \infty$ cannot be guaranteed in general, see \cite{li2010} for counter-examples and possible extensions. 

\begin{figure*}
\centering
\begin{subfigure}[b]{.47\textwidth}	
\centering
\includegraphics{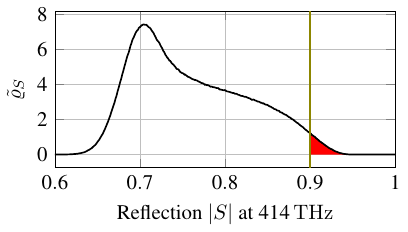}
\caption{Estimated \gls{pdf} of scattering parameter. Failure probability illustrated in red color.
}
\label{fig:kde}
\end{subfigure}
\hfill
\begin{subfigure}[b]{.47\textwidth}
\centering
\begin{tabular}{|c|c|c|c|}
\hline
$N^\text{MC}$ & $\mathbb{E}$ & $\sqrt{\mathbb{V}}$ & $\mathcal{F}$ \\
\hline 
\num{e3} & 0.7595 & 0.0661 & 2.20\,\% \\	\hline
\num{e4} & 0.7605 & 0.0658 & 1.85\,\% \\	\hline
\num{e5} & 0.7606 & 0.0660 & 2.04\,\% \\	\hline
\num{e6} & 0.7607 & 0.0660 & 2.07\,\% \\	\hline
\num{e7} & 0.7607 & 0.0660 & 2.06\,\% \\	\hline
\end{tabular}
\caption{Expectation, standard deviation and failure probability, i.e. $\mathcal F= P(|S|\ge 1- \alpha)$ for $\alpha=0.1$. Surrogate-based Monte Carlo estimation using $N^\text{MC}$ sample points.}
\label{tab:StatisticalMeasuresSF}
\end{subfigure}\\
\vspace{.5em}
\begin{subfigure}{1\textwidth}
\centering
\includegraphics{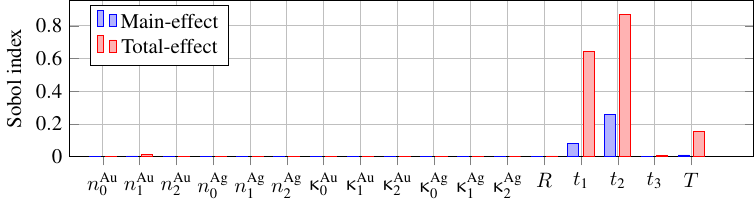}
\caption{Sobol indices computed with $\num{e5}$ sample points.}
\label{fig:sobol}
\end{subfigure}
\caption{\gls{pdf}, expectation, standard deviation, failure probability and Sobol indices for \num{17} beta-distributed input parameters.}
\end{figure*}

The \gls{pdf} $\varrho_S$ of $|S|$ is estimated by employing a kernel density estimator 
\begin{align}
\varrho_{S} \approx \tilde \varrho_{T}:=\frac{1}{hN^\text{MC}}\sum_{i=1}^{N^\text{MC}} K \left(\frac{T-|\tilde{S}^{(i)}|}{h}\right)
\end{align}
with $N^\text{MC}=\num{e7}$ samples, bandwidth $h=\num{e-3}$ and the Epanechnikov kernel \cite{epanechnikov1969}
\begin{align}
K(T) := \begin{cases}
\frac{3}{4} \left(1-{T}^2\right), ~~&{T}\in [-1,1],\\
0 , ~~&\mathrm{else}.
\end{cases}
\end{align}
The estimated expected values, standard deviations $\sqrt{\mathbb{V}}$ and failure probabilities for an increasing sample size $N^\text{MC}$ and $\alpha=0.1$ are given in Table~\ref{tab:StatisticalMeasuresSF}.
The estimated \gls{pdf} $\tilde \varrho_S$ is shown in Fig.~\ref{fig:kde}.

Sensitivity analysis is based on an analysis of variances (ANOVA) \cite{sobol2001}. The related metrics are commonly known as Sobol indices, where we will focus on the so-called main-effect (1st order) and total-effect (total order) indices, defined in \cite{homma1996importance}. 
In the context of the present work, estimations of the Sobol indices for the magnitude of the scattering parameter shall be based on sampling of the (mapped) polynomial approximation  $\tilde S:\Xi \rightarrow \mathbb C$. 
We use Saltelli's algorithm \cite{saltelli2002} with $N^{\text{sens}}=10^5$ sample points, resulting in $2(17+1)10^5=\num{3.6e6}$ surrogate model evaluations. 
The main-effect  and total-effect Sobol indices for each parameter are given in Fig.~\ref{fig:sobol}. 
The thickness of the dielectric layer $t_2$, the thickness of the upper gold layer $t_1$, the grating depth $T$ and the refractive index of the upper gold layer $n_1^{\text{Au}}$ are identified as the most sensitive parameters.
Moreover, since the sum of all main-effect sensitivity indices is approximately $33\%$, the remaining $67\%$ indicates higher order interactions, and thus strong coupling among the input parameters.

It is found that the considered model is highly sensitive to small geometrical variations. In particular, while geometrical variations in a range of only $\pm \SI{1.5}{nm}$ are considered, their impact is significantly higher than the one attributed to material uncertainty, which was modeled based on the measurement error provided by \cite{johnson1972}.

\section{Conclusion}
\label{sec:conclusion}

In this work we presented an efficient method to quantify uncertainties in the Maxwell source problem, assuming a moderately large number of input \gls{rv}s. Dimension adaptivity in combination with adjoint error correction and conformal maps is confirmed to be a promising technique to delay the curse-of-dimensionality. For the considered \gls{fe} model from nanoplasmonics, the comparison of the proposed adaptive algorithm with total degree gPC and isotropic Smolyak sparse grids shows significant gains in both accuracy and computational costs. In particular, with the adaptive scheme we were able to consider up to 17 parameters and achieve an accuracy of $\approx 10^{-3}$ with a few thousand numerical solutions of the deterministic model. 
To consider wider frequency ranges with possible poles in combination with large geometric uncertainties, a combination of polynomial and rational approximations is a topic of future research.

For the considered optical grating coupler, according to Sobol-sensitivity measures, geometrical parameters have been found to be the dominant source of input uncertainty. Although the modeling of their probability distributions could not be based on measurement data yet, this conclusion is substantiated by the very conservative choice of geometrical standard deviations.

\acknowledgements
The authors would like to thank R. Schuhmann for valuable discussions on the the topic of \gls{uq} in plasmonics and L. Scarabosio for
helpful discussions on sparse approximation. U. R\"omer and N. Georg acknowledge the funding by the Deutsche Forschungsgemeinschaft (DFG, German Research Foundation) – RO4937/1-1. The work of N. Georg, D. Loukrezis and S. Sch\"ops is also supported by the \textit{Excellence Initiative} of the German Federal and State Governments and the Graduate School of Computational Engineering at Technische Universit\"at Darmstadt. The work of D. Loukrezis is further supported by the BMBF via the research contract 05K19RDB.

\appendix

\section{Floquet boundary condition}
\label{sec:AppendixA}
To truncate the structure in the non-periodic direction at $\ensuremath{{\Gamma_{z^+}}}$, a Floquet absorbing boundary condition can be derived by splitting the electric field in the unbounded region $z\ge z^+$, where we assume vacuum permittivity $\varepsilon_0$ and vacuum permeability $\mu_0$, as 
\begin{equation}
\ensuremath{\mathbf{E}} = \ensuremath{\ensuremath{\mathbf{E}}^\mathrm{inc}} + \ensuremath{\ensuremath{\mathbf{E}}^\mathrm{sc}}, \label{eq:EincPlusEsc}
\end{equation}
where $\ensuremath{\ensuremath{\mathbf{E}}^\mathrm{inc}}$ and $\ensuremath{\ensuremath{\mathbf{E}}^\mathrm{sc}}$ represent the known incident field and the unknown scattered field, respectively.
As derived in \cite[Chapter 3]{bhattacharyya2006} and \cite[Chapter 12.2.1]{zhu2006}, the scattered field $\ensuremath{\ensuremath{\mathbf{E}}^\mathrm{sc}}$ can be represented as an infinite series of Floquet modes
\begin{align}
\ensuremath{\ensuremath{\mathbf{E}}^\mathrm{sc}} &= \sum_{\substack{m,n\in\mathbb{Z} \\ \ensuremath{\alpha} \in\{\text{TE, TM}\}}}c_{\ensuremath{\alpha},mn}\ensuremath{\mathbf{E}}_{\ensuremath{\alpha},mn}e^{-j\kappa_{mn}(z-z^+)},  \label{eq:floquetExpansion}\end{align}
where
\begin{align*}
\ensuremath{\mathbf{E}}_{\text{TE},mn} &\coloneqq \frac{e^{-j\left(k_{xm} x+k_{yn} y\right)} \left(k_{yn}\ensuremath{\ensuremath{\mathbf{e}}_x}  -k_{xm}\ensuremath{\ensuremath{\mathbf{e}}_y}\right)}{\sqrt{\ensuremath{{d_x}} \ensuremath{d_y}}\sqrt{k_{xm}^2+k_{yn}^2}}, \\
\ensuremath{\mathbf{E}}_{\text{TM},mn} &\coloneqq \frac{e^{-j\left(k_{xm} x+k_{yn} y\right)}\bigl(k_{xm}\ensuremath{\ensuremath{\mathbf{e}}_x}  +k_{yn} \ensuremath{\ensuremath{\mathbf{e}}_y} - \frac{k^2_{xm}+k^2_{yn}}{\kappa_{mn}}\ensuremath{\ensuremath{\mathbf{e}}_z}\bigr)}{\sqrt{\ensuremath{{d_x}} \ensuremath{d_y}}\sqrt{k_{xm}^2+k_{yn}^2}} , 
\end{align*}
with 
\begin{align}
k_{xm}\coloneqq \ensuremath{k^{\mathrm{inc}}_x}+\frac{2\pi m}{\ensuremath{{d_x}}},\quad  k_{yn} \coloneqq \ensuremath{k^{\mathrm{inc}}_y}+\frac{2\pi n}{\ensuremath{d_y}},\quad
\kappa_{mn} \coloneqq \sqrt{k_0^2-k_{xm}^2-k_{yn}^2}.
\end{align}
Thereby, we distinguish between \gls{te} modes $\ensuremath{\mathbf{E}}_{\text{TE},mn}$ and \gls{tm} modes $\ensuremath{\mathbf{E}}_{\text{TM},mn}$, fulfilling $\ensuremath{\mathbf{E}}\perp\ensuremath{\ensuremath{\mathbf{e}}_z}$ and $\ensuremath{\mathbf{H}}\perp \ensuremath{\ensuremath{\mathbf{e}}_z}$, respectively.
There exists only a finite number of propagating modes, i.e. $\kappa_{mn}\in \mathbb{R}$, depending on the wavenumber $k_0$, the angles of incidence $\ensuremath{\theta^\mathrm{inc}},~\ensuremath{\phi^\mathrm{inc}}$ and the dimensions $\ensuremath{{d_x}},~\ensuremath{d_y}$ of the unit cell. 

We introduce the operators $\ensuremath{\pi_\mathrm t} \left[\ensuremath{\mathbf{u}}\right]\coloneqq \ensuremath{\ensuremath{\mathbf{e}}_z} \times \ensuremath{\mathbf{u}}$ and $\ensuremath{\pi_\mathrm T} \left[\ensuremath{\mathbf{u}}\right] \coloneqq (\ensuremath{\ensuremath{\mathbf{e}}_z} \times \ensuremath{\mathbf{u}})\times \ensuremath{\ensuremath{\mathbf{e}}_z}$ such that 
\begin{align}
\ensuremath{\pi_\mathrm t} \left[ \ensuremath{\mathbf{H}}_{\ensuremath{\alpha},mn}e^{-j\kappa_{mn}(z-z^+)} \right] = 
\ensuremath{\pi_\mathrm t} \left[ \frac j{\omega \mu} \nabla\times \left(\ensuremath{\mathbf{E}}_{\ensuremath{\alpha},mn}e^{-j\kappa_{mn}(z-z^+)}\right)\right] = -Y_{\ensuremath{\alpha},mn} \ensuremath{\pi_\mathrm T}\left[\ensuremath{\mathbf{E}}_{\ensuremath{\alpha},mn}e^{-j\kappa_{mn}(z-z^+)}\right],\end{align}
with \begin{align}Y_{\ensuremath{\alpha},mn} &\coloneqq
\begin{cases}
\frac{\kappa_{mn}}{\omega \mu} &\text{for } \ensuremath{\alpha}=\text{TE},\nonumber\\
\frac{\omega \epsilon}{\kappa_{mn}} &\text{for } \ensuremath{\alpha}=\text{TM}.\nonumber 
\end{cases}
\end{align}
The incident plane wave $\ensuremath{\ensuremath{\mathbf{E}}^\mathrm{inc}}$ corresponds to the lowest order Floquet modes $\ensuremath{\mathbf{E}}_{\ensuremath{\alpha},00}$ with modal admittance $Y^\text{inc}$
\begin{align}
\ensuremath{\pi_\mathrm t} \left[ \ensuremath{\mathbf{H}}^\text{inc} \right] = Y^\text{inc} \ensuremath{\pi_\mathrm T}\left[\ensuremath{\ensuremath{\mathbf{E}}^\mathrm{inc}}\right], \quad Y^\text{inc} \coloneqq 
\begin{cases}
\frac{\sqrt{\epsilon}\cos(\ensuremath{\theta^\mathrm{inc}})}{\sqrt{\mu}} &\text{for}~\ensuremath{\alpha} = \text{TE},\\
\frac{\sqrt{\epsilon}}{\sqrt{\mu}\cos(\ensuremath{\theta^\mathrm{inc}})} &\text{for}~\ensuremath{\alpha} = \text{TM}.
\end{cases} 
\end{align}
By taking the cross product of the $\text{curl}$ of \eqref{eq:EincPlusEsc} with \ensuremath{\ensuremath{\mathbf{e}}_z}, the magnetic field above the structure is expressed as
\begin{align}
\ensuremath{\pi_\mathrm t} \left[ \ensuremath{\mathbf{H}} \right] + \sum_{\substack{m,n\in\mathbb{Z} \\ \ensuremath{\alpha} \in\{\text{TE, TM}\}}} \tilde c_{\ensuremath{\alpha},mn} Y_{\ensuremath{\alpha},mn}\ensuremath{\pi_\mathrm T}[\ensuremath{\mathbf{E}}_{\ensuremath{\alpha},mn}e^{-j\kappa_{mn}(z-z^+)}]= 2Y^\text{inc}\ensuremath{\pi_\mathrm T}\left[\ensuremath{\ensuremath{\mathbf{E}}^\mathrm{inc}}\right]. \label{eq:FloquetExact}
\end{align} For any $\ensuremath{\mathbf{u}},\ensuremath{\mathbf{v}} \in \left(L^2(\ensuremath{{\Gamma_{z^+}}})\right)^3$, the space of square-integrable complex vector functions on $\ensuremath{{\Gamma_{z^+}}}$, we introduce the inner product 
\begin{equation}
(\ensuremath{\mathbf{u}}, \ensuremath{\mathbf{v}})_\ensuremath{{\Gamma_{z^+}}} \coloneqq \int_{\ensuremath{{\Gamma_{z^+}}}} \ensuremath{\mathbf{u}} \cdot \ensuremath{\mathbf{v}}^* \mathop{}\!\mathrm{d} \ensuremath{\mathbf{x}}, 
\end{equation}
where the superscript $^*$ denotes complex conjugation.
Due to the orthogonality of the modal basis, i.e. 
\begin{subequations}
\begin{align}
\left(\ensuremath{\pi_\mathrm T}\left[\ensuremath{\mathbf{E}}_{\text{TE},mn}\right], \ensuremath{\pi_\mathrm T}\left[\ensuremath{\mathbf{E}}_{\text{TE},ij}\right]\right)_\ensuremath{{\Gamma_{z^+}}} &= \delta_{mi}\delta_{nj}\\ \left(\ensuremath{\pi_\mathrm T}\left[\ensuremath{\mathbf{E}}_{\text{TM},mn}\right], \ensuremath{\pi_\mathrm T}\left[\ensuremath{\mathbf{E}}_{\text{TM},ij}\right]\right)_\ensuremath{{\Gamma_{z^+}}} &= \delta_{mi}\delta_{nj}, \\ 
\left(\ensuremath{\pi_\mathrm T}\left[\ensuremath{\mathbf{E}}_{\text{TE},mn}\right], \ensuremath{\pi_\mathrm T}\left[\ensuremath{\mathbf{E}}_{\text{TM},ij}\right]\right)_\ensuremath{{\Gamma_{z^+}}} &= 0,
\end{align}
\end{subequations}
where $\delta$ denotes the Kronecker delta, the unknown coefficients $\tilde c_{\ensuremath{\alpha},mn}\in \mathbb{C}$ of the modal expansion \eqref{eq:FloquetExact} can be obtained as 
\begin{align}
\tilde c_{\ensuremath{\alpha},mn} = \left(\ensuremath{\pi_\mathrm T}\left[\ensuremath{\mathbf{E}}\right], \ensuremath{\pi_\mathrm T}\left[\ensuremath{\mathbf{E}}_{\ensuremath{\alpha},mn}\right]\right)_\ensuremath{{\Gamma_{z^+}}} = \left(\ensuremath{\pi_\mathrm T}\left[\ensuremath{\ensuremath{\mathbf{E}}^\mathrm{inc}}\right], \ensuremath{\pi_\mathrm T}\left[\ensuremath{\mathbf{E}}_{\ensuremath{\alpha},mn}\right]\right)_\ensuremath{{\Gamma_{z^+}}} + \underbrace{\left(\ensuremath{\pi_\mathrm T}\left[\ensuremath{\ensuremath{\mathbf{E}}^\mathrm{sc}}\right], \ensuremath{\pi_\mathrm T}\left[\ensuremath{\mathbf{E}}_{\ensuremath{\alpha},mn}\right]\right)_\ensuremath{{\Gamma_{z^+}}}}_{= c_{\ensuremath{\alpha},mn}}.
\label{eq:modal_coefficients}
\end{align}

Equation \eqref{eq:FloquetExact} represents the boundary condition to be imposed on $\ensuremath{{\Gamma_{z^+}}}$. 
In practice, the infinite sum of Floquet modes is truncated to \mbox{$-m_\text{max}\le m \le m_\text{max}$}, \mbox{$-n_\text{max}\le n \le n_\text{max}$}. 

In that case we obtain a boundary condition in the form of \eqref{eq:MaxwellBVP_Floquet} with 
\begin{align} 
\boldsymbol{\mathcal F}^\text{inc}  &= 2Y^\text{inc}\ensuremath{\pi_\mathrm T}\left[\ensuremath{\ensuremath{\mathbf{E}}^\mathrm{inc}}\right], \label{eq:FloquetHO} \\
\boldsymbol{ \mathcal G}(\ensuremath{\mathbf{E}}) &= \sum_{\substack{|m|\leq m_\text{max} \\ |n| \leq n_\text{max} \\ \ensuremath{\alpha} \in\{\text{TE, TM}\}}} \tilde c_{\ensuremath{\alpha},mn} Y_{\ensuremath{\alpha},mn}\ensuremath{\pi_\mathrm T}[\ensuremath{\mathbf{E}}_{\ensuremath{\alpha},mn}],  \label{eq:G_ho}
\end{align}
Further simplifications are possible if the dimensions of the unit cell are small enough, such that only the fundamental modes $\ensuremath{\mathbf{E}}_{\ensuremath{\alpha},00}$ propagate, and the boundary $\ensuremath{{\Gamma_{z^+}}}$ is placed sufficiently far away from the structure, such that all higher order modes are attenuated to a negligible amplitude. 
In this case, the fundamental mode is of particular interest and we may omit all evanescent higher order modes in \eqref{eq:G_ho}. 
In particular, we can employ the first-order absorbing boundary condition \cite[Chapter 13.4.1]{jin2015}, i.e. \eqref{eq:MaxwellBVP_Floquet} with 
\begin{align}
\boldsymbol{ \mathcal G}(\ensuremath{\mathbf{E}}) &=- \frac {\ensuremath{\ensuremath{\mathbf{k}}_\mathrm{t}^{\mathrm{inc}}}}{\omega \mu \ensuremath{k^{\mathrm{inc}}_z}} \left(  \ensuremath{\ensuremath{\mathbf{k}}_\mathrm{t}^{\mathrm{inc}}} \cdot \ensuremath{\pi_\mathrm T}\left[\ensuremath{\mathbf{E}}\right]\right) - \frac{\ensuremath{k^{\mathrm{inc}}_z}}{\omega \mu} \ensuremath{\pi_\mathrm T}\left[\ensuremath{\mathbf{E}}\right] \label{eq:FloquetFirstOrder},
\end{align} where $\ensuremath{\ensuremath{\mathbf{k}}_\mathrm{t}^{\mathrm{inc}}} \coloneqq \ensuremath{\pi_\mathrm T}[\ensuremath{\ensuremath{\mathbf{k}}^\mathrm{inc}}]$. 

The corresponding terms in the boundary conditions of the dual problem \eqref{eq:dual_strong} are given as
\begin{align}
\boldsymbol{\overline{\mathcal F}} =- \frac{j}{\omega\ensuremath{\mu_0}}\ensuremath{\pi_\mathrm T}\left[\ensuremath{\mathbf{E}}_{\ensuremath{\alpha},mn}\right],
\end{align}
and either
\begin{align}
\boldsymbol{ \overline{\mathcal G}} = -\sum_{\ensuremath{\alpha},m,n}\tilde d_{\ensuremath{\alpha},mn}^*Y_{\ensuremath{\alpha},mn}^*\ensuremath{\pi_\mathrm T}[\ensuremath{\mathbf{E}}_{\ensuremath{\alpha},mn}], 
\end{align} 
where $\tilde d_{\ensuremath{\alpha}, mn} = (\ensuremath{\pi_\mathrm T}[\ensuremath{\mathbf{E}}_{\ensuremath{\alpha},mn}], \ensuremath{\mathbf{z}}_\mathrm{T})_\ensuremath{{\Gamma_{z^+}}}$,
if \eqref{eq:G_ho} is used for the primal problem, or
\begin{align}
\boldsymbol{ \overline{\mathcal G}} = \frac {\ensuremath{\ensuremath{\mathbf{k}}_\mathrm{t}^{\mathrm{inc}}}}{\omega\ensuremath{\mu_0}\ensuremath{k^{\mathrm{inc}}_z}} \left(  \ensuremath{\ensuremath{\mathbf{k}}_\mathrm{t}^{\mathrm{inc}}} \cdot \ensuremath{\mathbf{z}}_\mathrm{T}\right) + \frac{\ensuremath{k^{\mathrm{inc}}_z}}{\omega \ensuremath{\mu_0}} \ensuremath{\mathbf{z}}_\mathrm{T}.
\end{align}
if lowest order Floquet boundary conditions \eqref{eq:FloquetFirstOrder} are employed in \eqref{eq:MaxwellBVP}.

\section{Details on FE discretization}
\label{sec:AppendixB}
The mesh is assumed to be periodic, i.e. the surface meshes on $\ensuremath{{\Gamma_{x^+}}}$ and $\ensuremath{{\Gamma_{x^-}}}$, as well as on $\ensuremath{{\Gamma_{y^+}}}$ and $\ensuremath{{\Gamma_{y^-}}}$, are respectively identical. Without loss of generality we further assume the vector of coefficients $\ensuremath{\mathbf{c}} \in \mathbb{C}^\ensuremath{{N_h}}$ 
to be ordered such that the boundary conditions imposed in $\eqref{eq:V}$ can be expressed as 
\renewcommand{\arraystretch}{1.2}
\begin{align*}
\ensuremath{\mathbf{c}} 
= \begin{bmatrix} 
\ensuremath{\mathbf{c}}_\text{inner} \\
\ensuremath{\mathbf{c}}_\ensuremath{{\Gamma_{z^+}}}\\
\ensuremath{\mathbf{c}}_\ensuremath{{\Gamma_{z^-}}}\\
\ensuremath{\mathbf{c}}_\ensuremath{{\Gamma_{x^+}}}\\
\ensuremath{\mathbf{c}}_\ensuremath{{\Gamma_{x^-}}}\\
\ensuremath{\mathbf{c}}_\ensuremath{{\Gamma_{y^+}}}\\
\ensuremath{\mathbf{c}}_\ensuremath{{\Gamma_{y^-}}}\\
\ensuremath{\mathbf{c}}_{\ensuremath{{\Gamma_{x^+}}}\cap\ensuremath{{\Gamma_{y^+}}}}\\
\ensuremath{\mathbf{c}}_{\ensuremath{{\Gamma_{x^-}}}\cap\ensuremath{{\Gamma_{y^+}}}}\\
\ensuremath{\mathbf{c}}_{\ensuremath{{\Gamma_{x^+}}}\cap\ensuremath{{\Gamma_{y^-}}}}\\
\ensuremath{\mathbf{c}}_{\ensuremath{{\Gamma_{x^-}}}\cap\ensuremath{{\Gamma_{y^-}}}}
\end{bmatrix} = \begin{bmatrix}
\mathbf{I} & 0 & 0 & 0 &0 \\
0 &   \mathbf{I} & 0 & 0 &0\\
0 & 0 & 0 & 0 & 0\\
0 & 0 & \mathbf{I} & 0 &0\\
0 & 0 & \mathbf{I}e^{-j\psi_x} & 0 & 0\\
0 & 0 & 0 & \mathbf{I} & 0\\
0 & 0 & 0 & \mathbf{I}e^{-j\psi_y} & 0\\
0 & 0 & 0 & 0 & \mathbf{I}\\
0 & 0 & 0 & 0 & \mathbf{I} e^{-j\psi_x}\\
0 & 0 & 0 & 0 & \mathbf{I} e^{-j\psi_y}\\
0 & 0 & 0 & 0 & \mathbf{I} e^{-j\left(\psi_x+\psi_y\right)}\\
\end{bmatrix} \begin{bmatrix} 
\ensuremath{\mathbf{c}}_\text{inner} \\
\ensuremath{\mathbf{c}}_\ensuremath{{\Gamma_{z^+}}}\\
\ensuremath{\mathbf{c}}_\ensuremath{{\Gamma_{x^+}}}\\
\ensuremath{\mathbf{c}}_\ensuremath{{\Gamma_{y^+}}}\\
\ensuremath{\mathbf{c}}_{\ensuremath{{\Gamma_{x^+}}}\cap\ensuremath{{\Gamma_{y^+}}}}
\end{bmatrix} = \mathbf{P} \ensuremath{\mathbf{c}}_\ensuremath{\mathrm{dof}},
\end{align*}
where we have introduced the reduced vector 
$
\ensuremath{\mathbf{c}}_\ensuremath{\mathrm{dof}} = \in \mathbb{C}^\ensuremath{{N_{\mathrm{DoF}}}}
$ of $\ensuremath{{N_{\mathrm{DoF}}}} < \ensuremath{{N_h}}$ \glspl{dof} and $\mathbf{I}$ denotes an identity matrix of appropriate size \cite[Chapter 13.1.2]{jin2015}.
\renewcommand{\arraystretch}{1}

Let $\mathbf{A}\in \mathbb C^{\ensuremath{{N_h}}\times \ensuremath{{N_h}}}$ and $\ensuremath{\mathbf{f}}\in \mathbb C^{\ensuremath{{N_h}}}$ be the system matrix and right-hand side vector, which are obtained by using \eqref{eq:nedelec} in \eqref{eq:WeakFormI}, as well as N{\'e}d{\'e}lec test functions. 
In case of using the higher-order Floquet port boundary condition, i.e. \eqref{eq:MaxwellBVP_Floquet} with \eqref{eq:FloquetHO}, the boundary integrals  lead to dense sub-blocks in the matrix $\mathbf{A}$, whereas \eqref{eq:FloquetFirstOrder} preserves the sparsity of the \gls{fe} matrix. 
The quasi-periodic and \gls{pec} boundary conditions \eqref{eq:V} on ansatz and test functions can be imposed conveniently using the matrix $\mathbf{P}\in \mathbb C^{\ensuremath{{N_h}}\times \ensuremath{{N_{\mathrm{DoF}}}}}$, leading to the reduced system
\begin{equation}
\mathbf{A}_{\mathrm{dof}}\mathbf{c}_{\mathrm{dof}}=\mathbf{P}^{\mathrm{H}} \mathbf{A} \mathbf{P} \ensuremath{\mathbf{c}}_\ensuremath{\mathrm{dof}}= \mathbf{P}^{\mathrm{H}} \ensuremath{\mathbf{f}} =\ensuremath{\mathbf{f}}_\ensuremath{\mathrm{dof}},\label{eq:DiscreteSystem}
\end{equation}
where $\mathbf P^{\mathrm{H}}$ denotes the Hermitian transpose of $\mathbf P$. 
Functions spanned by the reduced DoF form a proper subspace of $\eqref{eq:V}$.

\section{Convergence study of gPC and sparse-grid projection}
\label{sec:AppendixC}
\begin{figure}
\centering
\includegraphics{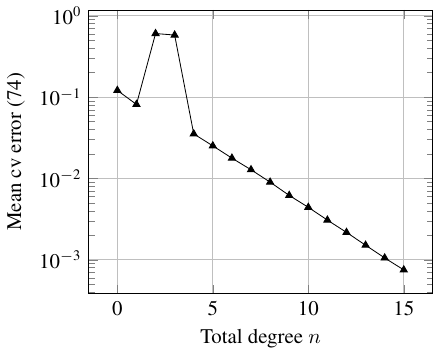}
\caption{Convergence study of a \gls{gpc} approximation using pseudo-spectral projection and sparse Gauss quadrature.}
\label{fig:2dConvSparse}
\end{figure}

Since the computational cost for a proper convergence study in the 17-dimensional setting is too high, we restrict us again to the two most sensitive parameters, i.e. $t_1, t_2$, and repeat the gPC convergence study with sparse Gaussian quadrature. Again, we use a random cross-validation sample of size $N^\text{MC}=1000$ to compute the error \eqref{eq:mean_error}. Results are presented in Fig.~\ref{fig:2dConvSparse}. It can be observed that, similar to Fig.~\ref{fig:sf_conv}, the error slightly increases up to order $3$ before a convergent behavior can indeed be observed. 

\begin{thebibliography}{10}

\bibitem{chkifa2015breaking}
Chkifa, A., Cohen, A., and Schwab, C., Breaking the curse of dimensionality in
sparse polynomial approximation of parametric pdes, {\em Journal de
Math{\'e}matiques Pures et Appliqu{\'e}es}, 103(2):400--428, 2015.

\bibitem{scarabosio2016}
Scarabosio, L., Shape uncertainty quantification for scattering transmission
problems, PhD thesis, ETH Zurich, 2016.

\bibitem{xiu2009}
Xiu, D., Fast numerical methods for stochastic computations: A review, {\em
Commun. Comput. Phys.}, 5(2-4):242--272, 2 2009.

\bibitem{ghanem1991}
Ghanem, R.G. and Spanos, P.D., {\em Stochastic Finite Elements: A Spectral
Approach}, Springer, New York, 1991.

\bibitem{babuska2004}
Babu{\v{s}}ka, I., Tempone, R., and Zouraris, G.E., Galerkin finite element
approximations of stochastic elliptic partial differential equations, {\em
SIAM J. Numer. Anal.}, 42(2):800--825, 2004.

\bibitem{babuska2010}
Babu{\v{s}}ka, I., Nobile, F., and Tempone, R., A stochastic collocation method
for elliptic partial differential equations with random input data, {\em SIAM
Review}, 2:317--355, 2010.

\bibitem{xiu2002}
Xiu, D. and Karniadakis, G.E., The {Wiener-Askey} polynomial chaos for
stochastic differential equations, {\em SIAM J. Sci. Comput.},
24(2):619--644, 2002.

\bibitem{xiu2005}
Xiu, D. and Hesthaven, J.S., High-order collocation methods for differential
equations with random inputs, {\em SIAM J. Sci. Comput.}, 27(3):1118--1139,
2005.

\bibitem{chkifa2014}
Chkifa, A., Cohen, A., and Schwab, C., High-dimensional adaptive sparse
polynomial interpolation and applications to parametric {PDE}s, {\em Found.
Comput. Math.}, 14(4):601--633, 2014.

\bibitem{gerstner2003}
Gerstner, T. and Griebel, M., Dimension-adaptive tensor-product quadrature,
{\em Computing}, 71(1):65--87, 2003.

\bibitem{narayan2014}
Narayan, A. and Jakeman, J.D., Adaptive {L}eja sparse grid constructions for
stochastic collocation and high-dimensional approximation, {\em SIAM J. Sci.
Comput.}, 36(6), 2014.

\bibitem{nobile2008a}
Nobile, F., Tempone, R., and Webster, C.G., An anisotropic sparse grid
stochastic collocation method for partial differential equations with random
input data, {\em SIAM J. Numer. Anal.}, 46(5):2411--2442, 2008.

\bibitem{ernst2018}
Ernst, O.G., Sprungk, B., and Tamellini, L., Convergence of sparse collocation
for functions of countably many gaussian random variables (with application
to elliptic {PDE}s), {\em SIAM J. Numer. Anal.}, 56(2):877--905, 2018.

\bibitem{benner2015}
Benner, P. and Schneider, J., Uncertainty quantification for {M}axwell's
equations using stochastic collocation and model order reduction, {\em Int.
J. Uncertain. Quantif.}, 5(3), 2015.

\bibitem{benner2015survey}
Benner, P., Gugercin, S., and Willcox, K., A survey of projection-based model
reduction methods for parametric dynamical systems, {\em SIAM review},
57(4):483--531, 2015.

\bibitem{bodendiek2014adaptive}
Bodendiek, A. and Bollh{\"o}fer, M., Adaptive-order rational arnoldi-type
methods in computational electromagnetism, {\em BIT Numerical Mathematics},
54(2):357--380, 2014.

\bibitem{bonizzoni2018pade}
{Bonizzoni}, F., {Nobile}, F., {Perugia}, I., and {Pradovera}, D.,
{Least-Squares Pad{\'e} approximation of parametric and stochastic Helmholtz
maps}, {\em ArXiv e-prints}, May 2018.

\bibitem{chantrasmi2009pade}
Chantrasmi, T., Doostan, A., and Iaccarino, G., Pad{\'e}--{L}egendre
approximants for uncertainty analysis with discontinuous response surfaces,
{\em J. Comput. Phys.}, 228(19):7159--7180, 2009.

\bibitem{scarabosio2019multilevel}
Scarabosio, L., Multilevel monte carlo on a high-dimensional parameter space
for transmission problems with geometric uncertainties, {\em International
Journal for Uncertainty Quantification}, 9(6), 2019.

\bibitem{silva2017}
Silva-Oelker, G., Aylwin, R., Jerez-Hanckes, C., and Fay, P., Quantifying the
impact of random surface perturbations on reflective gratings, {\em IEEE
Trans. Antennas Propag.}, 66(2):838--847, 2017.

\bibitem{doelz2019}
{D{\"o}lz}, J., {A higher order perturbation approach for electromagnetic
scattering problems on random domains}, {\em arXiv e-prints}, p.
arXiv:1907.05501, Jul 2019.

\bibitem{jantsch2018sparse}
Jantsch, P. and Webster, C.
\newblock Sparse grid quadrature rules based on conformal mappings.
\newblock In {\em Sparse Grids and Applications-Miami 2016}, pp. 117--134.
Springer, 2018.

\bibitem{trefethen2013}
Trefethen, L.N., {\em Approximation theory and approximation practice}, Vol.
128, SIAM, 2013.

\bibitem{jakeman2015}
Jakeman, J.D. and Wildey, T., Enhancing adaptive sparse grid approximations and
improving refinement strategies using adjoint-based a posteriori error
estimates, {\em J. Comput. Phys.}, 280:54--71, 2015.

\bibitem{butler2013propagation}
Butler, T., Dawson, C., and Wildey, T., Propagation of uncertainties using
improved surrogate models, {\em SIAM/ASA J. Uncertain.}, 1(1):164--191, 2013.

\bibitem{loukrezis2019assessing}
{Loukrezis}, D., {R{\"o}mer}, U., and {De Gersem}, H., Assessing the
performance of {L}eja and {C}lenshaw-{C}urtis collocation for computational
electromagnetics with random input data, {\em Int. J. for Uncertain.
Quantif.}, 9(1):33--57, 2019.

\bibitem{loukrezis2019approximation}
{Loukrezis}, D. and {De Gersem}, H., Approximation and uncertainty
quantification of stochastic systems with arbitrary input distributions using
weighted {L}eja interpolation, {\em arXiv e-prints}, p. arXiv:1904.07709, Apr
2019.

\bibitem{farcas2019multilevel}
{Farcas}, I.G., {Latz}, J., {Ullmann}, E., {Neckel}, T., and {Bungartz}, H.J.,
{Multilevel adaptive sparse {L}eja approximations for {B}ayesian inverse
problems}, {\em arXiv e-prints}, p. arXiv:1904.12204, Apr 2019.

\bibitem{Bos2019}
van~den Bos, L., Sanderse, B., Bierbooms, W., and van Bussel, G., Bayesian
model calibration with interpolating polynomials based on adaptively weighted
{Leja} nodes, {\em Communications in Computational Physics}, 27(1):33--69,
2019.

\bibitem{loukrezis2019adaptive}
Loukrezis, D. and De~Gersem, H., Adaptive sparse polynomial chaos expansions
via {Leja} interpolation, {\em arXiv preprint arXiv:1911.08312}, 2019.

\bibitem{genet2007}
Genet, C. and Ebbesen, T.W., Light in tiny holes, {\em Nature}, 445:39, jan
2007.

\bibitem{preiner2008}
Preiner, M.J., Shimizu, K.T., White, J.S., and Melosh, N.A., Efficient optical
coupling into metal-insulator-metal plasmon modes with subwavelength
diffraction gratings, {\em Appl. Phys. Lett.}, 92(11):113109, 2008.

\bibitem{pitelet2019influence}
Pitelet, A., Schmitt, N., Loukrezis, D., Scheid, C., Gersem, H.D., Cirac\`{i},
C., Centeno, E., and Moreau, A., Influence of spatial dispersion on surface
plasmons, nanoparticles, and grating couplers, {\em J. Opt. Soc. Am. B},
36(11):2989--2999, Nov 2019.

\bibitem{schmitt2019optimization}
Schmitt, N., Georg, N., Bri\`{e}re, G., Loukrezis, D., H\'{e}ron, S., Lanteri,
S., Klitis, C., Sorel, M., R\"{o}mer, U., Gersem, H.D., V\'{e}zian, S., and
Genevet, P., Optimization and uncertainty quantification of gradient index
metasurfaces, {\em Opt. Mater. Express}, 9(2):892--910, Feb 2019.

\bibitem{loukrezis2019robust}
Loukrezis, D., Galetzka, A., and De~Gersem, H., Robust adaptive least squares
polynomial chaos expansions in high-frequency applications, {\em arXiv
preprint arXiv:1912.07725}, 2019.

\bibitem{weng2015}
Weng, T.W., Zhang, Z., Su, Z., Marzouk, Y., Melloni, A., and Daniel, L.,
Uncertainty quantification of silicon photonic devices with correlated and
non-{Gaussian} random parameters, {\em Opt. express}, 23(4):4242--4254, 2015.

\bibitem{hiptmair2018}
Hiptmair, R., Scarabosio, L., Schillings, C., and Schwab, C., Large deformation
shape uncertainty quantification in acoustic scattering, {\em Adv. Comput.
Math.}, pp. 1--44, 2018.

\bibitem{babuska2007}
Babu{\v{s}}ka, I., Nobile, F., and Tempone, R., A stochastic collocation method
for elliptic partial differential equations with random input data, {\em SIAM
J. Numer. Anal.}, 45(3):1005--1034, 2007.

\bibitem{jankoski2019stochastic}
Jankoski, R., R{\"o}mer, U., and Sch{\"o}ps, S., Stochastic modeling of
magnetic hysteretic properties by using multivariate random fields, {\em Int.
J. Uncertain. Quantif.}, 9(1), 2019.

\bibitem{lebrun2009rosenblatt}
Lebrun, R. and Dutfoy, A., Do {R}osenblatt and {N}ataf isoprobabilistic
transformations really differ?, {\em Probabilist. Eng. Mech.},
24(4):577--584, 2009.

\bibitem{lemaitre2010}
Le~Maitre, O.P. and Knio, O.M., {\em Spectral Methods for Uncertainty
Quantification: With Applications to Computational Fluid Dynamics},
Scientific Computation, Springer Netherlands, 2010.

\bibitem{xiu2010}
Xiu, D., {\em Numerical Methods for Stochastic Computations: A Spectral Method
Approach}, Princeton University Press, Princeton, 2010.

\bibitem{barthelmann2000}
Barthelmann, V., Novak, E., and Ritter, K., High dimensional polynomial
interpolation on sparse grids, {\em Adv. Comput. Math.}, 12(4):273--288,
2000.

\bibitem{bungartz2004}
Bungartz, H.J. and Griebel, M., Sparse grids, {\em Acta Numer.}, 13:147--269,
2004.

\bibitem{klimke2005}
Klimke, A. and Wohlmuth, B.I., Algorithm 847: Spinterp: Piecewise multilinear
hierarchical sparse grid interpolation in {MATLAB}, {\em ACM Trans. Math.
Softw.}, 31(4):561--579, 2005.

\bibitem{nobile2008}
Nobile, F., Tempone, R., and Webster, C.G., A sparse grid stochastic
collocation method for partial differential equations with random input data,
{\em SIAM J. Numer. Anal.}, 46(5):2309--2345, 2008.

\bibitem{schieche2012}
Schieche, B., {\em Unsteady Adaptive Stochastic Collocation on Sparse Grids},
PhD Thesis, TU Darmstadt, 2012.

\bibitem{hale2008}
Hale, N. and Trefethen, L.N., New quadrature formulas from conformal maps, {\em
{SIAM} J. Numer. Anal.}, 46(2):930--948, 2008.

\bibitem{hale2009}
Hale, N., On the use of conformal maps to speed up numerical computations, PhD
thesis, Oxford University, 2009.

\bibitem{kosloff1993}
Kosloff, D. and Tal-Ezer, H., A modified {C}hebyshev pseudospectral method with
an {$\mathcal O (N-1)$} time step restriction, {\em J. Comput. Phys.},
104(2):457--469, 1993.

\bibitem{Boyd_2001aa}
Boyd, J.P., {\em {Chebyshev} and {Fourier} Spectral Methods}, Dover
Publications, 2 edition, 2001.

\bibitem{berrut2004}
Berrut, J. and Trefethen, L., Barycentric {L}agrange interpolation, {\em SIAM
Review}, 46(3):501--517, 2004.

\bibitem{smolyak1963}
Smolyak, S.A., Quadrature and interpolation formulas for tensor products of
certain classes of functions, {\em Dokl. Acad. Nauk SSSR}, 4:240--243, 1963.

\bibitem{becker2001optimal}
Becker, R. and Rannacher, R., An optimal control approach to a posteriori error
estimation in finite element methods, {\em Acta Numer.}, 10:1--102, 2001.

\bibitem{butler2012posteriori}
Butler, T., Constantine, P., and Wildey, T., A posteriori error analysis of
parameterized linear systems using spectral methods, {\em SIAM J. Matrix
Anal. Appl.}, 33(1):195--209, 2012.

\bibitem{Roemer2015}
R\"omer, U. and Sch\"ops, S., Adjoint error estimation for a pseudo-spectral
approach to stochastic field-circuit coupled problems, In {\em Proc. Appl.
Math. Mech.)}, Vol.~15, pp. 711--714. Wiley-VCH, October 2015.

\bibitem{teckentrup2013}
Teckentrup, A.L., Scheichl, R., Giles, M.B., and Ullmann, E., Further analysis
of multilevel {M}onte {C}arlo methods for elliptic {PDE}s with random
coefficients, {\em Numer. Math.}, 125(3):569--600, 2013.

\bibitem{jin2015}
Jin, J.M., {\em The Finite Element Method in Electromagnetics}, John Wiley \&
Sons, Hoboken, 2015.

\bibitem{monk2003}
Monk, P., {\em Finite element methods for {M}axwell's equations}, Oxford
University Press, 2003.

\bibitem{nedelec1980}
Nedelec, J.C., Mixed finite elements in {$R^3$}, {\em Numer. Math.},
35(3):315--341, 1980.

\bibitem{Hiptmair_2002aa}
Hiptmair, R., Finite elements in computational electromagnetism, {\em Acta
Numer.}, 11:237--339, 2002.

\bibitem{aylwin2019domain}
Aylwin, R., Jerez-Hanckes, C., Schwab, C., and Zech, J., Domain uncertainty
quantification in computational electromagnetics, {\em SAM Research Report},
2019, 2019.

\bibitem{cstTutorial}
{CST AG}.
\newblock Optical applications with {CST Microwave Studio}, 2012.
\newblock
\url{https://www.cst.com/content/events/downloads/euc2012/talk_5-3-1_cst_euc_2012.pdf}
(Accessed: 2018-03-12).

\bibitem{johnson1972}
Johnson, P.B. and Christy, R.W., Optical constants of the noble metals, {\em
Physical review B}, 6(12):4370, 1972.

\bibitem{maier2007}
Maier, S.A., {\em Plasmonics: fundamentals and applications}, Springer Science
\& Business Media, New York, 2007.

\bibitem{geuzaine2009}
Geuzaine, C. and Remacle, J.F., Gmsh: A {3-D} finite element mesh generator
with built-in pre-and post-processing facilities, {\em Int. J. Numer. Meth.
Eng.}, 79(11):1309--1331, 2009.

\bibitem{alnaes2015}
Aln{\ae}s, M., Blechta, J., Hake, J., Johansson, A., Kehlet, B., Logg, A.,
Richardson, C., Ring, J., Rognes, M.E., and Wells, G.N., The {FEniCS} project
version 1.5, {\em Archive of Numerical Software}, 3(100):9--23, 2015.

\bibitem{cst}
{CST AG}.
\newblock {CST} {STUDIO} {SUITE} 2016, 2018.

\bibitem{braibant1984}
Braibant, V. and Fleury, C., Shape optimal design using {B}-splines, {\em
Comput. Methods Appl. Mech. Eng.}, 44(3):247--267, 1984.

\bibitem{piegl1997}
Piegl, L. and Tiller, W., {\em The {NURBS} Book}, Springer, 2 edition, 1997.

\bibitem{lopes2011}
Lopes, R.H.
\newblock Kolmogorov-{S}mirnov test.
\newblock In {\em International Encyclopedia of Statistical Science}, pp.
718--720. Springer, 2011.

\bibitem{back2011}
B\"ack, J., Nobile, F., Tamellini, L., and Tempone, R.
\newblock Stochastic spectral {G}alerkin and collocation methods for {PDE}s
with random coefficients: a numerical comparison.
\newblock In: Hesthaven, J. and Ronquist, E. (Eds.), {\em Spectral and High
Order Methods for Partial Differential Equations}, Vol.~76 of Lecture Notes
in Computational Science and Engineering, pp. 43--62. Springer, 2011.
\newblock Selected papers from the ICOSAHOM '09 conference, June 22-26,
Trondheim, Norway.

\bibitem{feinberg2015}
Feinberg, J. and Langtangen, H.P., Chaospy: An open source tool for designing
methods of uncertainty quantification, {\em J. Comput. Science}, 11:46--57,
2015.

\bibitem{li2010}
Li, J. and Xiu, D., Evaluation of failure probability via surrogate models,
{\em J. Comput. Phys.}, 229(23):8966--8980, 2010.

\bibitem{epanechnikov1969}
Epanechnikov, V.A., Non-parametric estimation of a multivariate probability
density, {\em Theor. Probab. Appl.}, 14(1):153--158, 1969.

\bibitem{sobol2001}
Sobol, I.M., Global sensitivity indices for nonlinear mathematical models and
their {Monte Carlo} estimates, {\em Math. Comput. Simul.}, 55(1):271 -- 280,
2001.

\bibitem{homma1996importance}
Homma, T. and Saltelli, A., Importance measures in global sensitivity analysis
of nonlinear models, {\em Reliab. Eng. Syst. Safe.}, 52(1):1--17, 1996.

\bibitem{saltelli2002}
Saltelli, A., Making best use of model evaluations to compute sensitivity
indices, {\em Comput. Phys. Commun.}, 145(2):280--297, 2002.

\bibitem{bhattacharyya2006}
Bhattacharyya, A.K., {\em Phased array antennas: Floquet analysis, synthesis,
BFNs and active array systems}, Vol. 179, John Wiley \& Sons, Hoboken, 2006.

\bibitem{zhu2006}
Zhu, Y. and Cangellaris, A.C., {\em Multigrid finite element methods for
electromagnetic field modeling}, Vol.~28, John Wiley \& Sons, Hoboken, 2006.

\end{thebibliography}
\end{document}